\documentclass[11pt]{article}
\usepackage[utf8]{inputenc}
\usepackage{verbatim}
\if{ 
\usepackage{draftwatermark}
\SetWatermarkText{Draft - Please do not distribute}
\SetWatermarkScale{1}
\SetWatermarkLightness{0.85}
}\fi

\usepackage{authblk}

\title{A model for COVID-19 with isolation, quarantine and testing as control measures}

\author[1]{M.S. Aronna\thanks{soledad.aronna@fgv.br}}
\author[1]{R. Guglielmi\thanks{roberto.guglielmi@fgv.br}}
\author[1]{L.M. Moschen\thanks{Lucas.MachadoMoschen@gmail.com}}

\affil{Escola de Matem\'atica Aplicada, FGV EMAp - Rio de Janeiro, Brazil}

\date{\today}

\usepackage{tikz}
\usepackage{relsize}    
\usetikzlibrary{decorations.pathreplacing} 
\usetikzlibrary{shapes,arrows}

\usepackage{bm}

\usepackage{graphicx}
\usepackage{amssymb,amsmath}
\usepackage{color}

\usepackage[hyphens]{url}
\usepackage[breaklinks]{hyperref}
\hypersetup{breaklinks=true}
\usepackage{breakurl}

\usepackage{caption}
\usepackage{stackrel}
\usepackage{float}

\usepackage{caption,subcaption}
\def\R{\mathbb{R}}

\newtheorem{theorem}{Theorem}[section]

\newtheorem{remark}[theorem]{Remark}

\newtheorem{proposition}[theorem]{Proposition}

\usepackage{geometry}
\begin{document}

\maketitle

\begin{abstract}
    In this article we propose a compartmental model for the dynamics of Coronavirus Disease 2019 (COVID-19). We take into account the presence of asymptomatic infections and the  main policies that have been adopted so far to contain the epidemic: isolation (or social distancing) of a portion of the population, quarantine for confirmed cases and testing. We model isolation by separating the population in two groups: one composed by key-workers that keep working during the pandemic and have a usual contact rate, and a second group consisting of people that are enforced/recommended to stay at home. We refer to quarantine as strict isolation, and it is applied to confirmed infected cases. 
    
    In the proposed model, the proportion of people in isolation, the level of contact reduction and the testing rate are control parameters that can vary in time, representing policies that evolve in different stages. 
    We obtain an explicit expression for the basic reproduction number $\mathcal{R}_0$ in terms of the parameters of the disease and of  the control policies. In this way we can quantify the effect that isolation and testing have in the evolution of the epidemic.
    We present a series of simulations to illustrate different realistic scenarios.
    From the expression of  $\mathcal{R}_0$ and the simulations we conclude that isolation (social distancing) and testing among asymptomatic cases are fundamental actions to control the epidemic, {and the stricter these measures are and the sooner they are implemented,} the more lives can be saved. Additionally, we show that people that remain in isolation significantly reduce their probability of contagion, so risk groups should be recommended to  maintain a low contact rate during the course of the epidemic.
\end{abstract}

\section{Introduction}

In late December 2019, several cases of an {\em unknown pneumonia} were identified in the city of Wuhan, Hubei province, China \cite{outbreakncipcdcchina}. Some doctors of Wuhan conjectured that it could be severe acute respiratory syndrome (SARS) cases \cite{stephaniehegarty2020}. Many of the found cases had visited or were related to the Huanan Seafood Wholesale Market.
On 31 December 2019, the World Health Organization (WHO) China Country Office was informed of these cases of pneumonia detected in Wuhan City and, up to 3 January 2020, a total of 44 patients with this unknown pneumonia were reported to WHO \cite{whopneumoniachina}. 
In the beginning of January 2020 Chinese officials  ruled out the hypothesis that the cases were of SARS \cite{aljazeeratimiline2020}, and a few days later the cause was identified to be a new coronavirus that was named  SARS-CoV-2. The name given to the  infectious disease caused by SARS-CoV-2 is COVID-19. 

\vspace{5pt}

The first death due to COVID-19 was reported on 9 January and it was a 61-year-old man in Wuhan \cite{newyorkqina2020}.
After mid January infected cases were  reported in Thailand, Japan, Republic of Korea, and other provinces in China \cite{outbreakncipcdcchina}. On 22 January the Chinese authorities announced the quarantine of greater Wuhan. From that time on the virus rapidly spread in many Asiatic countries, reached Europe and the United States. 
On 28 February, with more than 80.000 confirmed cases and nearly 3.000 deaths globally, WHO  increased the assessment of the risk of spread and risk of impact of COVID-19 to very high at the global level \cite{whoreport392020}. 
On 9 March 2020, with nearly 400 deaths, Italian government ordered the total lock-down of the national territory 
\cite{whoreport492020}.
And a few days later, on March 11, WHO declared that COVID-19 was characterized as a pandemic \cite{whoreport512020}.

\vspace{5pt}

In March 2020 several nations across the five continents closed their borders, declared forced isolation for the whole population except for essential workers and/or imposed strict measures of social distancing. Detailed information on the actions taken by each country can be found in the report \cite{OxCGRT}.
At that time WHO recommended, apart from social distancing measures, that it was essential to test intensively \cite{whoremarks16media2020}.
The indications were to test every suspected case, to isolate till recovery any positive individual, and to track and test all contacts in the past two days of new confirmed cases. 

\vspace{5pt}

 The most common symptoms of COVID-19 are fever, cough and shortness of breath. Most of the cases result in mild or no symptoms, but some progress to viral pneumonia and multi-organ failure. At this moment, it is estimate that 1 out of 5 cases needs hospitalization \cite{whocoronavirus2020}. It is yet difficult to estimate the mortality of this virus. Mortality depends on one side on early detection and appropriate treatment, but the rate itself can only be calculated if the real number of infected people is known. There is enough evidence to assure that a significant portion of the infections is asymptomatic \cite{lavezzo2020suppression,mizumoto2020estimating,nishiura2020estimation}, which makes it difficult to detect them and thus to calculate the effective mortality of COVID-19.  WHO, by March 2020, estimated a death rate of 3,4\% worldwide \cite{whoremarks3media2020}. But in some countries, as Italy, France, Spain and UK, the rate between deaths and confirmed cases up to May 2020 is higher than 10\% \cite{worldometers-countries}.

\vspace{5pt}

In this article we propose a compartmental model for the dynamics of COVID-19. We take into account the presence of asymptomatic infections, and also the main policies that have been adopted by several countries in the past months to fight this disease, {these being}: isolation, quarantine and testing. We model isolation by separating the population in two groups: one composed by {\em key-workers} that keep working during the pandemic and having a usual contact rate, and the other group consisting of people that are enforced/recommended to stay at home. Certainly, in the group of people that maintain a high contact rate one can also include people that do not respect social distancing restrictions, that has lately shown to be significant in some countries.
We refer to quarantine as strict isolation, and it is applied to confirmed infected cases. Testing is supposed to be applied to all symptomatic cases, and to a portion of the population selected using some of the criteria adopted by health organizations (see {\em e.g.} \cite{investigationgovuk,guidanceCDC}). The idea to analyze the quantitative effects of non-pharmaceutical interventions, such as isolation and social distancing, on the evolution of the epidemic was inspired by the work~\cite{ferguson2020impact}.
 
 \vspace{5pt}

 For the proposed model, we obtain an expression for the basic reproduction number $\mathcal{R}_0$ in terms of the parameters of the disease and of  the control parameters. In this way we can quantify the effects that isolation and testing have on the epidemic.
We exhibit a series of simulations to illustrate different realistic situations. We compare, in particular, different levels of isolation and testing.
From the expression of  $\mathcal{R}_0$ and the simulations, we conclude that isolation (social distancing) and testing among asymptomatic cases are fundamental actions to control the epidemic, {and the stricter these measures are and the sooner they are implemented,} the more lives can be saved. {Additionally, we show that people that remain in isolation significantly reduce their probability of contagion, so risk groups should be recommended to  maintain a low contact rate during the course of the epidemic.}

 \vspace{5pt}

Several mathematical models for COVID-19 have appeared recently in the literature. At the time being, the flux of publications is very high, so it is difficult to keep track of everything that is being published. We next mention and describe some of the models more closely related to ours. In \cite{casella2020can} they consider a simple model, with infected and reported infected compartments, and they assume that the transmission rate $\beta$ is a function of a control $u,$ this is $\beta = \beta(u).$ They analyze feedback control strategies, where the control depends on the number of reported cases. In \cite{djidjou2020optimal} they consider mild and severe cases, the latter having a reduced transmission rate since they are assumed to be in isolation. They use a time-dependent control $c$ of reduction of contacts for the whole population, and optimize with respect to this control. An SEIR model with quarantine for suspected and infected cases cases is considered in \cite{shi2020seir}, and in \cite{liu2020predicting} they take into account unreported cases, asymptomatic individuals and quarantine for identified cases.



 \vspace{5pt}


The article is organized as follows.  In Section \ref{sec:model} we introduce the model, and we discuss its structure. In Section \ref{sec:R0} we show an expression of the basic reproduction number $\mathcal{R}_0$ in terms of the parameters and we propose an equivalent threshold. Estimation of realistic parameters and numerical simulations are given in Section \ref{sec:NumSim}, while Section \ref{sec:conclusions} is devoted to the conclusions and a description of possible continuations of this research. Finally, in the Appendix we include the analytical computations of the expression of $\mathcal{R}_0$ and a sensitivity analysis with respect to the involved parameters.

\section{The model description}\label{sec:model}

We set up a model to describe the spread of the virus SARS-CoV-2 through a susceptible population. Building upon a usual SEIR model, we obtain a more structured {one}, which is tailored on the {current} experience of the COVID-19 epidemic, and which also allows to convey the effects of the non-pharmaceutical intervention policies {being} adopted by several countries to face its outbreak.

First of all, we normalize {to $1$} the total population of $N$ individuals, so that all the compartments (and their sub-compartments) introduced below represent the proportion of individuals of the total population in such compartment. We will assume the population remains constant over time (i.e., we neglect the natural birth and death rates).
We start by splitting the population in the compartments listed in Table~\ref{tab:compartment}.

\begin{table}[h]
\centering
\begin{tabular}{|c|c|}
\hline
 {\bf Compartment} & {\bf Description} \\[0.5ex]
\hline
$S$ & susceptible\\[0.3ex]
\hline
    $E$ & exposed\\[0.3ex]
\hline
    $I$ &  infectious \\
\hline
$A$ & asymptomatic and infectious\\
\hline
$Q$ & infected in quarantine (including hospitalized) \\
\hline
$R$ & recovered\\
\hline
\end{tabular}
\caption{\label{tab:compartment}List of aggregated compartments}
\end{table}

More specifically, the compartment $S$ collects all the individuals that are susceptible to the virus. Once an individual from $S$ gets exposed to the virus, moves to the compartment $E$. Let us point out that individuals in $E$, though already exposed to the virus, are not contagious yet. After a given \emph{latent time}, an individual in $E$ becomes infectious, and thus is allocated to the compartment $I$. At this stage, after a suitable time, the individual may either remain infectious but asymptomatic (or with mild symptoms), in which case moves to the compartment $A$, or may show clear symptoms onset, thus being tested and then quarantined either at home or at the hospital, and being assigned to the compartment $Q$. Finally, individuals in $A$ and $Q$ will eventually be removed from those compartments and will end up
{{either} in the compartment $R$ after a \emph{recovery time} or  {dead}}.

We will assume that the fraction of asymptomatic individuals among all infected is given by a certain probability $\alpha\in (0,1)$. It is intuitive that the presence of a relevant portion of asymptomatic infectious individuals plays a major role in the spread of the epidemic, as observed in the current outbreak~\cite{mizumoto2020estimating,nishiura2020estimation}. Indeed, an asymptomatic individual will maintain a high contact rate, and thus might infect more susceptible individuals with respect to an infectious individual with symptoms {that is in} quarantine. In our model, we always refer to the effective contact rate $\beta$, which is given by the product between the \emph{transmissibility} $\nu$ (i.e., probability of infection given contact between a susceptible and infected individual), and the \emph{average rate of contact} {$c$} between susceptible and infected individuals. In Tables~\ref{Tab:ParPathogen} and \ref{Tab:ParPubPolicy} we list all the parameters of the model and their description.

The model as described so far takes into account several characteristics of the pathogen and its spread in a susceptible population. We now want to add further structural features to the model in order to include the non-pharmaceutical interventions adopted by public policies to contain the epidemic. In particular, we assume the following conditions.
\begin{itemize}
    \item[i)] A part $p$ of the population is in  isolation (either voluntarily, or as a result of public safety policies). The remaining $1-p$ of the population instead gathers all those so-called ``key workers" (such as physicians and paramedicals, workers in logistics and distribution, food production, security, and others), that  must continue with a regular activity, thus maintaining a large contact rate and being exposed to a higher risk of infection. We will generically refer to such $1-p$ part of population as the \emph{active} population, as opposed to the population in isolation. In this group we can also include people that simply do not respect social distancing, and thus maintain a high contact rate. A situation like this has been observed in countries were monitoring was not strict and a significant percentage of the population did not respect isolation.
    \item[ii)] The fraction $1-p$ of active population has an effective contact rate $\beta$, whereas the $p$ part of population in isolation has a contact rate reduced by a factor $r$, thus its compound contact rate is $r\beta$. We will therefore refer to such portion of the population as in {\em $r$-isolation}.
    \item[iii)]  A centralized controller (such as the national health system) may intervene on the system by testing a portion of the population to check for the infectious pathogen. We assume the testing kit to be reliable, that is, we neglect the possibility of false positive/negative. As a rule, then, an individual from the compartment $S$ will always test negative, an individual from $I$ or $A$ always positive, while an individual from $E$ will result positive with a probability $\delta\in [0,1]$. In this way, even though the individuals in $E$ are not contagious, we account for the possibility that they might result positive to the test, depending on the stage of development of the pathogen in that specific individual and to the efficacy of the testing kit.
\end{itemize}

Let us notice that, in general, the effective contact rate $\beta$ depends on a variety of factors, including the density of population in a given country/region. However, during a pandemic, even the effective contact rate of the individuals not in isolation may be reduced by increased awareness (for example, maintaining the social distancing), or by respecting stricter safety protocols and by availability of proper Personal Protection Equipment (PPE), including face shields, masks, gloves, soap, and so on.

According to the above description, each compartment $S$, $E$, $I$ and $A$ is partitioned as follows: $S = S_f \cup S_r$, where $S_f$ are susceptible and active, while $S_r$ are susceptible and in $r$-isolation; $E = E_f \cup E_r$, where $E_f$ are exposed and active, while $E_r$ are exposed and in $r$-isolation; $I = I_f \cup I_{r}$, where $I_f$ are infectious and active, $I_{r}$ are infectious and in $r$-isolation; $A = A_f \cup A_{r}$, where $A_f$ are asymptomatic infectious and active, $A_{r}$ are asymptomatic infectious and in $r$-isolation. The compartment $Q$ collects all the infected individuals who have been tested positive, either after onset of severe symptoms, or because of a sample test among the population, according to the procedure described in iii) of the above list. Let us stress that, among these compartments, only the individuals in $Q$ are aware of being infected, and thus contagious, hence they are either hospitalized or at home, but in both cases they follow strict procedures to reduce their contact rate to $0$. Finally, we will use the compartments $R$ for the recovered and immune individuals, and $D$ for the disease-induced deaths. Both these last compartments will be removed from the dynamics and will end up in the counter system~\eqref{eq:counters}.  Moreover, we point out that the portion $p$ of the population in $r$-isolation is predetermined at the initial time of the evolution, reflecting the public policy in place in that specific period of time. Of course, such fraction $p$ may be updated at a later time, accordingly to newer (stricter or looser) public policies.


\bigskip

The first set of constants, related to the pathogen itself (assuming no mutation occurs in the time of epidemic, or if so, the mutation does not affect such parameters of the virus) and its induced disease, are collected in Table~\ref{Tab:ParPathogen}. A graphical representation of the course of the disease for symptomatic carriers can be seen in Figure \ref{timeline}.

\begin{table}[ht]
\centering
\begin{tabular}{|c|c|}
\hline
 {\bf Par.} & {\bf Description} \\[0.5ex]
\hline
  $ \tau$  & inverse of the latent time from exposure to infectiousness onset\\[0.3ex]
\hline
    $\sigma$ & 
inverse of the time from infectiousness onset to possible symptoms onset
    \\[0.3ex]
\hline
     $\theta$ & inverse of mean incubation time (i.e. $\theta^{-1} = \tau^{-1} + \sigma^{-1} $)
     \\[0.3ex]
\hline
    $\alpha$ & proportion of asymptomatic infections\\[0.3ex]
\hline
    $\gamma_1$ & recovery rate for asymptomatic or mild symptomatic cases\\[0.3ex]
\hline
    $\gamma_2$ & recovery rate for severe and critical cases  \\[0.3ex]
\hline
    $\mu$ & mortality rate among confirmed cases\\[0.3ex]
\hline
    $\delta$ & probability of detection by testing in compartment $E$ \\
\hline
\end{tabular}
\caption{Parameters of {COVID-19}}
\label{Tab:ParPathogen}
\end{table}

\begin{figure}[!h]
    \centering
    \label{timeline}
    \begin{tikzpicture}[scale=1]
    \draw[ultra thick, ->] (0,0) -- (14,0);
    
    \foreach \x in {1,5,9,13}
    \draw (\x cm,3pt) -- (\x cm,-3pt);
    
    \draw[ultra thick] (1,0) node[below=3pt,thick] {exposure} node[above=3pt] {};
    \draw[ultra thick] (5,0) node[below=3pt,thick, text width=2.5cm,align=center] {infectiousness \\ onset} node[above=3pt] {};
    \draw[ultra thick] (9,0) node[below=3pt, thick, text width=2cm,align=center] {symptoms \\ onset} node[above=3pt] {};
    \draw[ultra thick] (13,0) node[below=3pt] {recovery} node[above=3pt] {};
    
    \draw[ultra thick] (3,0) node[above=3pt,thick] {$\tau^{-1}$} node[below=3pt] {};
    \draw[ultra thick] (7,0) node[above=3pt,thick, text width=2.5cm,align=center] {$\sigma^{-1}$} node[below=3pt] {};
    \draw[ultra thick] (11,0) node[above=3pt, thick, text width=2cm,align=center] {$\gamma_2^{-1}$} node[below=3pt] {};
    \draw [black, ultra thick ,decorate,decoration={brace,amplitude=5pt}] (1,0.7)  -- (9,0.7) 
    node [black,midway,above=5pt,thick] {incubation period: $\theta^{-1}$};
    
    \draw [thick,->, red, dotted] (12,0) -- (12,-1) node[below=1pt] {death};

\end{tikzpicture}
    \caption{Disease timeline for symptomatic cases}
    \label{timeline}
\end{figure}
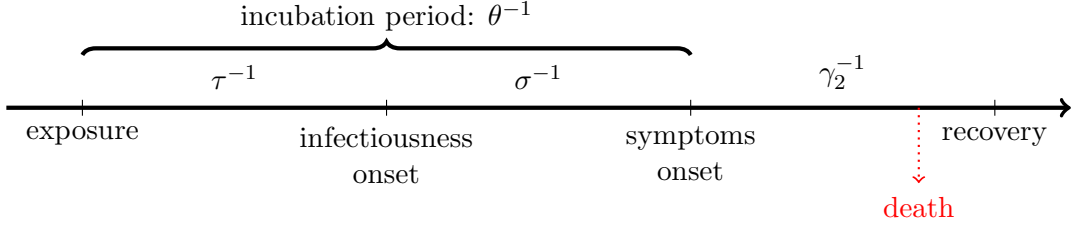

The second set of parameters is related to public policies, and consists of the parameters in Table~\ref{Tab:ParPubPolicy}.
{Let us recall at this point that $\beta$ varies in each territory, depending mainly on population density and behaviour.}
These constants may be used as control parameters, via the tuned lockdown as decided by the public policies (reflecting on $p$ and partially on $r$), the awareness of the population in respecting the social distancing among individuals and in the widespread use of personal protection equipment (expressed by $\beta$ and partially by $r$), the availability of testing kits, that results in a higher or lower value of $\rho$. 

\begin{table}[h]
\centering
\begin{tabular}{|c|c|}
\hline
 {\bf Par.} & {\bf Description} \\[0.5ex]
\hline
$\beta(t)$ & transmission rate at time $t$ (proportional to contact rate)\\[0.3ex]
\hline
    $r(t)$ & reduction coefficient of transmission rate\\
    & for people in isolation at time $t$\\[0.3ex]
\hline
    $\rho(t)$ & testing rate of people with mild or no symptoms at time $t$ \\
\hline
$p(t)$ & \shortstack{proportion of the population in $r$-isolation} \\
\hline
\end{tabular}
\caption{Parameters of Public Policies interventions}
\label{Tab:ParPubPolicy}
\end{table}

The extended state variable of the system thus becomes
\[
\tilde{X} = (E_f,E_r,I_f,I_{r},A_f,A_r,Q,S_f,S_r,R,D)\, ,
\]
where the description of the compartments is given in Table \ref{table:compartments}.

\begin{figure}[h!]
    \begin{center}
    \resizebox{400pt}{!}{
        \begin{tikzpicture}[squarednode/.style={rectangle, minimum size=15mm, thick},
            fontscale/.style = {font=\relsize{#1}}]
            \node[squarednode, draw = black, fill = blue!70] (nodeSf) at (0,0) [fontscale = 2] {$\boldsymbol{S_f}$};
            \node[squarednode, draw = black, fill = blue!50] (nodeSr) at (0,-4) [fontscale = 2] {$\boldsymbol{S_r}$};
            \node[squarednode, draw = black, fill = orange] (nodeEf) at (3,0) [fontscale = 2] {$\boldsymbol{E_f}$};
            \node[squarednode, draw = black, fill = orange!60] (nodeEr) at (3,-4) [fontscale = 2] {$\boldsymbol{E_r}$};
            \node[squarednode, draw = black, fill = orange] (nodeIf) at (6,0) [fontscale = 2] {$\boldsymbol{I_f}$};
            \node[squarednode, draw = black, fill = orange!60] (nodeIr) at (6,-4) [fontscale = 2] {$\boldsymbol{I_r}$};
            \node[squarednode, draw = black, fill = orange] (nodeAf) at (9,1) [fontscale = 2] {$\boldsymbol{A_f}$};
            \node[squarednode, draw = black, fill = orange!60] (nodeAr) at (9,-5) [fontscale = 2] {$\boldsymbol{A_r}$};
            \node[squarednode, draw = black, fill = yellow] (nodeQ) at (11,-2) [fontscale = 2] {$\boldsymbol{Q}$};        
            \node[squarednode, draw = black, fill = red] (nodeD) at (14,-4) [fontscale = 2] {$\boldsymbol{D}$};
            \node[squarednode, draw = black, fill = green!80] (nodeR) at (14,0) [fontscale = 2] {$\boldsymbol{R}$};

            \draw[->] (-3,-2) -- node[midway,above,sloped] {$1 - p$} (nodeSf.west);
            \draw[->] (-3,-2) -- node[midway,below,sloped] {$p$} (nodeSr.west);
            \draw[->] (nodeSf) -- node[midway,above,sloped] {$\beta$} (nodeEf);
            \draw[->] (nodeEf) -- node[midway,above,sloped] {$\tau$} (nodeIf);
            \draw[->] (nodeIf) -- node[midway,above,sloped] {$\alpha\sigma$} (nodeAf);
            \draw[->] (nodeIf) -- node[midway,above,sloped] {$(1 - \alpha)\sigma$} (nodeQ);
            \draw[->] (nodeAf) -- node[midway,above,sloped] {$\gamma_1$} (nodeR);
            \draw[->] (nodeQ) -- node[midway,above,sloped] {$\gamma_2$} (nodeR);
            \draw[->] (nodeSr) -- node[midway,below,sloped] {$\beta r$} (nodeEr);
            \draw[->] (nodeEr) -- node[midway,below,sloped] {$\tau$} (nodeIr);
            \draw[->] (nodeIr) -- node[midway,below,sloped] {$\alpha\sigma$} (nodeAr);
            \draw[->] (nodeIr) -- node[midway,below,sloped] {$(1 - \alpha)\sigma$} (nodeQ);
            \draw[->] (nodeQ) --  node[midway,below] {$\mu$} (nodeD);
            \draw[->] (nodeEf.south) to[bend right=10] node[pos = 0.5, above] {$\delta\rho$} (10.21,-1.88);
            \draw[->] (nodeEr.north) to[bend right=-10] node[pos = 0.5, below] {$\delta\rho$} (10.21,-2.12);
            \draw[->] (nodeAf) -- node[midway,above] {$\rho$} (nodeQ);
            \draw[->] (nodeAr) --  node[midway,below] {$\rho$} (nodeQ);
            \draw[->] (nodeAr) to[bend right=10]  node[pos=0.7,below,sloped] {$\gamma_1$} (nodeR.south);

        \end{tikzpicture}}
    \end{center}
    \caption{Model diagram}
    \label{diagram}
\end{figure}
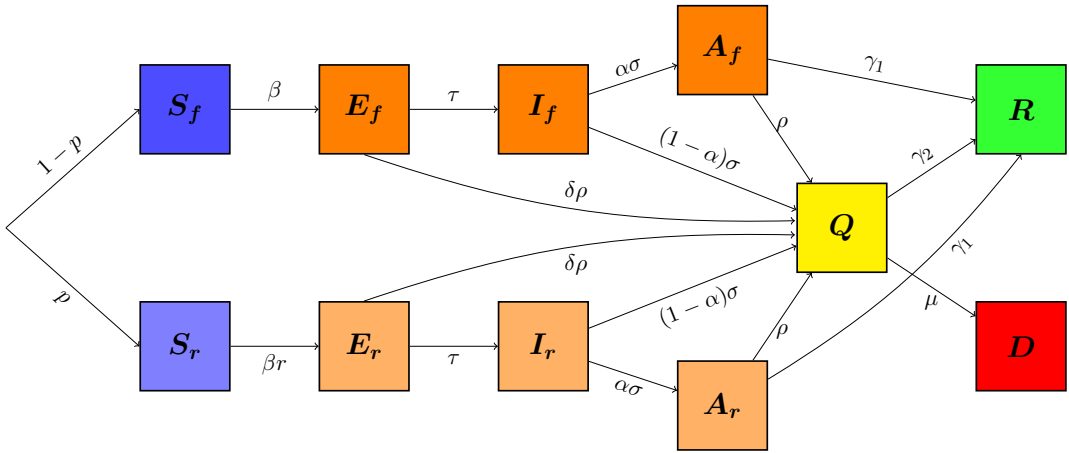

\begin{table}[h]
\begin{center}
\begin{tabular}{|c|c|}
\hline
 {\bf Compartment} & {\bf Description} \\[0.5ex]
\hline
    $E_f$ & exposed, not in isolation, not contagious\\[0.3ex]
\hline
   $E_r$ & exposed, in $r$-isolation, not contagious\\[0.3ex]
\hline
    $I_f$ & infected and contagious, not in isolation \\
\hline
$I_r$ & infected and contagious, in $r$-isolation \\
\hline
$A_f$ & asymptomatic and contagious, not in isolation\\
\hline
$A_r$ & asymptomatic and contagious, in $r$-isolation\\
\hline
$Q$ & infected and tested positive, in enforced quarantine \\
\hline
$S_f$ & susceptible not in isolation\\[0.3ex]
\hline
$S_r$ & susceptible in $r$-isolation\\[0.3ex]
\hline
$R$ & recovered and immune\\
\hline
$D$ & dead\\
\hline
\end{tabular}
\caption{List of extended compartments}
\label{table:compartments}
\end{center}
\end{table}
We focus in particular on the evolution of the variable 
$$
X = (E_f,E_r,I_f,I_{r},A_f,A_r,Q,S_f,S_r)\, ,
$$
that follows the model
\begin{equation}\label{eq:SEIRwQ}
\begin{array}{l}
\Dot{E}_f = \beta(t) S_f \left[I_f + A_f + r(t)(I_{r} + A_r)\right] -\rho(t) \delta E_f- \tau E_f \\[0.5ex]
\Dot{E}_r = r(t) \beta(t)  S_r \left[I_f + A_f + r(t)(I_{r} + A_r)\right] -\rho(t)\delta E_r - \tau E_r\\[0.5ex]
\Dot{I}_f = \tau E_f - \sigma I_f - \rho(t) I_f \\[0.5ex]
\Dot{I}_{r} = \tau E_r - \sigma I_{r} - \rho(t) I_{r} \\[0.5ex]
\Dot{A}_f = \sigma\alpha I_f - \rho(t) A_f - \gamma_1 A_f \\[0.5ex]
\Dot{A}_{r} = \sigma\alpha I_{r} - \rho(t) A_{r} - \gamma_1 A_r \\[0.5ex]
\Dot{Q} = \sigma (1-\alpha) (I_f + I_{r}) + \rho(t) \big[\delta(E_f+E_r)+I_f + I_{r} + A_f + A_{r}\big] - \gamma_2 Q - \mu Q \\[0.5ex]
\Dot{S}_f = -\beta(t) S_f [I_f + A_f + r(t) ( I_{r} + A_r)] \\[0.5ex]
\Dot{S}_r = -r(t) \beta(t)  S_r [I_f + A_f + r(t) ( I_{r} + A_r)]
\end{array}
\end{equation}
while the evolution of the states
\begin{equation}\label{eq:counters}
\left.
\begin{array}{l}
\Dot{R} = \gamma_1 (A_f + A_{r}) + \gamma_2 Q\\[0.5ex]
\Dot{D} = \mu Q
\end{array}
\right.
\end{equation}
only provides counters for the proportion (over the total population) of recovered and dead individuals, respectively.
 See the compartmental diagram associated to this model in Figure \ref{diagram}.

\begin{remark}[About the testing rate $\rho$]\label{RemarkTests}
The parameter $\rho$ indicates the proportion of the population  presenting either mild or no symptoms that is tested daily. It can also be thought as the inverse of the mean duration that an infected person passes without being tested. For instance, if the system manages to detect, each day, 5\% of the asymptomatic infections, then $\rho = 0.05.$ If we are in an ideal ``trace and test'' situation (see e.g. South Korea \cite{guardian2020}), in which for each confirmed infection, his/her recent contacts are rapidly and efficiently traced and tested, then $\rho$ will be greater and this will have an impact in the {\em basic reproduction number} (see Section \ref{sec:R0}).
\end{remark}

Recalling that testing is supposed to be applied, at least, to all sufficiently symptomatic cases, we add a counter for the positive tests $T(t)$ until time $t,$ which evolves according to the equation
\begin{equation*}
\dot T = \sigma(1-\alpha)(I_f+I_{r}) + \rho(t) \big(\delta(E_f+E_r)+I_f+I_{r}+A_f+A_{r}\big)\; .
\end{equation*}
Having this quantity, one can estimate the total number of tests in each territory using the {\em testing positive rate} of that location, which is the ratio between reported cases and tests done \cite{whoreport331509,worldometers-testing}.

\begin{remark}[About symptoms and quarantine]
In our framework, we assume that all cases with sufficiently severe symptoms are (tested and) quarantined, and we set the parameter $\alpha \in (0,1)$ to be the fraction of asymptomatic cases, including the cases with mild symptoms. But we can adapt our model to a scenario in which even severe symptoms are not tested until critical. In this case, with a very large value for $\alpha$, only a small portion among the symptomatic individuals enters directly to $Q,$ while the others need to be tested (according to the sampling testing $\rho$ among the population) to be quarantined.
\end{remark}

System~\eqref{eq:SEIRwQ} is endowed with the set of initial conditions given by the vector
\begin{equation}\label{eq:ICb}
X_0 = (E_{f,0},E_{r,0},I_{f,0},I_{r,0},A_{f,0},A_{r,0},Q_{0},S_{f,0},S_{r,0})
\end{equation}
{with components in the interval $[0,1]$}.
Setting
$$
\mathcal{C}_1
:= \big\{X = (x_i)_{i = 1,\ldots,9} \in \R^9 : x_i\in [0,1],\, \text{ for } i=1,\dots 9\big\}\; ,
$$
the cube of states with entries between $0$ and $1$, it is easy to check that $\mathcal{C}_1$ is invariant under the flow of system~\eqref{eq:SEIRwQ}, that is, given an initial condition $X_0\in \mathcal{C}_1$, the solution $X(t)$ to~\eqref{eq:SEIRwQ}-\eqref{eq:ICb} remains in $\mathcal{C}_1$ for all $t>0$.

\begin{remark}[Possible extensions of the model]
We collect here some va\-ri\-ations of model~\eqref{eq:SEIRwQ} that can be formulated in the same framework considered in this paper.
\begin{enumerate}
    \item One might consider a small but not negligible contact rate between susceptible individuals and people in the compartment $Q$, accounting for infections (mainly of medicals and paramedicals) occurred during hospitalization of an infected individual, or for individuals tested positive in enforced quarantine at home, which do not comply strictly to the isolation procedures and end up infecting relatives or other contacts. In this case, the equations for the evolution of the susceptible compartments shall be completed with additional terms involving $\varepsilon$ in the following way:
\begin{equation*}
\begin{array}{l}
\Dot{S}_f = -\beta(t) S_f [I_f + A_f + r(t) ( I_{r} + A_r) {\ +\  \varepsilon Q}] \, ,\\[0.7ex]
\Dot{S}_r = -r(t) \beta(t)  S_r [I_f + A_f + r(t) ( I_{r} + A_r) {\ +\  \varepsilon Q}]\, ,
\end{array}
\end{equation*}
and the same terms with opposite sign shall appear in the equation corresponding to $\Dot{Q}$.
    \item At the current stage, it is still not clear how long the immunity of a recovered individual lasts, with a number of findings tending towards a rather long immunization period~\cite{An2020recovered,time-reinfection,wajnberg2020humoral}. For this reason, in~\eqref{eq:SEIRwQ} we assume that a recovered individual will remain immune over the time framework considered in the different scenarios. However, the model can easily describe the case of recovered individuals becoming susceptible again, by adding a transfer term from the compartment $R$ to $S_f$ and $S_r$, with a coefficient depending on the inverse of the average immunization period. Similarly, the model can include the case of reactivation of the virus in an individual previously declared recovered (and not newly exposed to the virus), by inserting a transfer term from the compartment $R$ into $I_f$ and $I_r$, with appropriate coefficients depending on the probability of the reactivation of the virus and on the inverse of the average time of reactivation. However, at the moment there are not strong evidences supporting such possibility~\cite{time-reinfection}.
    \item A crucial issue while coping with the outbreak of the epidemic, which leads to the so-called urge of \emph{flattening the curve}, is whether the number of critical cases in need of intensive care (IC) treatment (due to respiratory failure, shock, and multiple organ dysfunction or failure) would saturate the number of available intensive care units (ICUs).\\
    This parameter can be estimated directly from model~\eqref{eq:SEIRwQ}, considering for each country the number of available ICUs and the percentage of positive confirmed cases requiring IC treatment. For example, this percentage has been estimated to be about $6\%$ for China~\cite{WHO-Ch}, and up to $12\%$ for Italy~\cite{GrassPeseCecco2020,Remuzzi2020}. As an alternative, it would be possible to insert a further compartment $C$ in model~\eqref{eq:SEIRwQ} counting the number of the individuals needing ICU treatment, by modifying the equations corresponding to the compartments $Q$ and $D$ as follows:
\begin{equation*}
        \begin{split}
  \Dot{Q} &= \sigma (1-\alpha) I + \rho(t) \big[\delta E + I + A\big] - \gamma_2 Q {\ -\ \tau_c Q}\\
         {\dot C } &= {\tau_c Q\  -\  \mu_c C\  -\  \gamma_c C}\\
            \dot D &= {\mu_c C}
        \end{split}
\end{equation*}
with suitable coefficients $\tau_c$, $\mu_c$ and $\gamma_c$ denoting the inverse of the time from symptoms onset to critical symptoms, the mortality of critical cases, and the recovery rate for critical cases, respectively.
    \item In this paper we have considered the whole population as a fixed number of individuals during the time period of the evolution. It is of course possible to consider the case of an evolving total population, by including in the model~\eqref{eq:SEIRwQ} the natural birth and mortality rate. In particular, newborns of susceptible individuals shall enter the corresponding susceptible compartment, whereas it is not clear whether the offspring of an infectious individual would be infectious, in such case the newborn shall move directly to the compartment $I$. On the other hand, the natural mortality rate shall act on each compartment of  system~\eqref{eq:SEIRwQ}, as well as on $R$ in system~\eqref{eq:counters}.
\end{enumerate}
\end{remark}

\section{The basic reproduction number for model~\eqref{eq:SEIRwQ}}\label{sec:R0}

We are interested in determining the basic reproduction number $\mathcal{R}_0$ associated with system~\eqref{eq:SEIRwQ}. To do this, we assume to fix a time interval $[t_0,t_1]$ such that the coefficients $\beta(t)$, $r(t)$ and $\rho(t)$ are constant over $[t_0,t_1]$. This is coherent with the setting of the scenarios simulated in  Section~\ref{subsec:Scenarios}, where we assume such coefficients to be piecewise constant functions, sharing the same switching times, {that represent different phases of restrictions and policies}. Thus, according {to the calculations given in} the Appendix~\ref{App1} and the parameters in Tables~\ref{Tab:ParPathogen} and~\ref{Tab:ParPubPolicy}, we obtain that the value of $\mathcal{R}_0$ for each time interval between two consecutive switching times is given by
\begin{equation}\label{R0wheneps=0}
\mathcal{R}_0 = \frac{1}{2}\left(
\varphi + \sqrt{\varphi^2 + \frac{4\sigma\alpha}{\rho + \gamma_1}\varphi}\;
\right)\, ,
\end{equation}
with
\begin{equation}
\label{varphi}
    \varphi = \frac{\beta \tau (1 - (1 - r^2) p)}{(\rho\delta + \tau)(\sigma + \rho)}\ .
\end{equation}
From this explicit formula for the reproduction number $\mathcal{R}_0$, we can highlight the qualitative dependence of $\mathcal{R}_0$ on each parameter of the system, in particular:
\begin{itemize}
\item[-] If the effective contact rate $\beta$ increases, then $\mathcal{R}_0$ increases.
\item[-] Focusing on the coefficient $1 - (1-r^2)p$, we realize that closer is $p$ to $1$, and smaller is $r$, that is, as larger is the portion of population in $r$-isolation and as stricter is the reduction factor $r$ in its contact rate, lower $\mathcal{R}_0$ becomes.
\item[-] If $\alpha$ increases, that is, if there is a larger proportion of asymptomatic infectious individuals, then $\mathcal{R}_0$ increases.
\item[-] If $\sigma$ increases, corresponding to shorter onset time, then $\mathcal{R}_0$ decreases.
\item[-] If either $\rho$ or $\gamma_1$ increase, i.e., either the control action by testing is strengthened, for example through an improved tracing and tracking system, or the recovery rate improves, for example, because of new and more effective treatments, then $\mathcal{R}_0$ decreases.
\item[-] If $\delta$ increases, for example, as a result of improved testing kits able to detect the infection at an earlier stage, then $\mathcal{R}_0$ decreases.
\end{itemize}
Moreover, we can characterize the crucial condition $\mathcal{R}_0\le 1$ by means of a simpler expression than~\eqref{R0wheneps=0}, as described in the next result.

\begin{proposition}[Alternative threshold]
\label{PropR0}
Set
$$
\mathcal{T}_0 := \frac{\beta \tau [1 - (1 - r^2) p]}{(\rho\delta + \tau)(\sigma + \rho)} \left(
1 + \frac{\sigma\alpha}{\rho + \gamma_1}
\right)\, .
$$
{Then} $\mathcal{R}_0$ is {smaller} than (respectively, equal to or greater than) 1 if and only if the same relation holds for $\mathcal{T}_0$. In particular, if $\varphi > 1$ (see \eqref{varphi}), then $\mathcal{R}_0 > 1$ and $\mathcal{T}_0 > 1$.
\end{proposition}

We postpone the proof of Proposition~\ref{PropR0} {to} the Appendix~\ref{App2}. A more quantitative analysis of the dependence of the threshold $\mathcal{T}_0$ on the parameters of the model is developed in the Appendix~\ref{App3}.

\begin{remark}[{About no testing among asymptomatic carriers}]\label{rem:R0tau}
If we consider the case of $\rho = 0$, that is, the situation without sample testing among the asymptomatic population, then the basic reproduction number $\mathcal{R}_0$ is independent of the latent time $\tau$. {In particular, $\mathcal{T}_0$ becomes  $$\frac{\beta  [1 - (1 - r^2) p]}{\sigma} \left(
1 + \frac{\sigma\alpha}{\gamma_1}
\right).$$}\; 
\end{remark}

\begin{remark}[On the time-dependent reproduction number]
\label{rem:timeDepR0}
{Relation } \eqref{R0wheneps=0} gives {an expression of} $\mathcal{R}_0$, that is the reproduction number  in a totally susceptible population. As the epidemic evolves, a portion of the population becomes immune to the disease, and this makes the reproduction number decrease. More precisely, when $p=0$ and all the population has the same contact rate, the time-dependent reproduction number is given by $\mathcal{R}(t) =  S(t) \mathcal{R}_0$, where $S(t)$ is the susceptible portion of the population. In our model, since the groups of active individuals and in $r$-isolation evolve differently (see Scenario A$_4$ and Figure \ref{A4comparison} below), the time-dependent reproduction number $\mathcal{R}(t)$  is given by the formula \eqref{R0wheneps=0} where
$\varphi$ in \eqref{varphi} is 
$$
\frac{\beta \tau [{S_f(t)} + r^2 S_r(t)]}{(\rho\delta + \tau)(\sigma + \rho)}\; .
$$
We do not take into account this time-variation of the reproduction number in our numerical results, since we are only interested in the value of the reproduction number at the beginning of each phase, where  $S(t)$ is close to 1.
\end{remark}

\begin{remark}[On herd immunity]
\label{rem:herd}
 Herd immunity is defined as the proportion of the population that needs to be immunized in order to naturally slow down the spread of the disease. It depends on the value of the basic reproduction number in the following way: herd immunity level equals $1 - \dfrac{1}{\mathcal{R}_0}.$ So the bigger $\mathcal{R}_0$, the higher the herd immunity. 
In connection with above Remark \ref{rem:timeDepR0}, we highlight that herd immunity is achieved at the time $t$ when $\mathcal{R}(t)$ equals 1.
\end{remark}

\section{Numerical simulations}\label{sec:NumSim}

\subsection{Retrieving parameters}

In Table \ref{parameter_values} we collect some parameter values estimated in the literature, in order to do realistic simulations. Recall the description of the parameters given in Tables \ref{Tab:ParPathogen}-\ref{Tab:ParPubPolicy}.

\begin{table}[ht]
\centering
\begin{tabular}{|c|c|c|c|}
\hline
 {\bf Par.} & {\bf Value - Range} & {\bf Reference} & {\bf Remark} \\[0.5ex]
\hline
{$\beta$} &  0.7676&  \cite{shen2020modelling} & \ref{transmission} \\
\hline
$\tau^{-1}$ &   $\tau^{-1} = \theta^{-1} - \sigma^{-1}$ & \cite{liang2020impacts} & \ref{latent}\\
\hline
$\sigma^{-1}$ &  \shortstack{1 - 3 days} & \cite{Read_etal,zhang2020evolving,Zhou2020} & \ref{rk:infectiouswindow} \\
\hline
$\theta^{-1}$ & 5.1 - 6.4 days & \cite{backer2020incubation,lauer2020incubation,liang2020impacts} & \ref{incubationWuhan}\\
\hline
$\gamma_1$  & 7.5 - 12 days & \cite{WHO-Ch,Hu_etal_Asy} & \ref{recovery1} \\
\hline
$\gamma_2$ & 
\begin{tabular}{c}
15 - 22 days
\end{tabular}
 & \cite{WHO-Ch,Zhou2020} & \ref{recovery2} \\
\hline
$\mu$  & [0.03/14,0.1/14] &  \cite{whoremarks3media2020,wang2020updated} & \ref{mu}\\
\hline
$\alpha$  & [0.265,0.643] & \cite{wiki:diamond,lavezzo2020suppression} & \ref{asymp} \\
\hline
$p$  & $[0 , 1]$ & \cite{OxCGRT} & \ref{pr}  \\
\hline
$r$  & $[0 , 1]$  & \cite{OxCGRT} & \ref{pr}\\
\hline
$\rho$ & [0,0.5] & \cite{worldometers-testing} & \ref{rho} \\
\hline
$\delta$ & 1 & \cite{HarvardMedicine} & \ref{delta}\\
\hline
\end{tabular}
\caption{Realistic range of parameters values}
\label{parameter_values}
\end{table}

\vspace{5pt}

Several remarks regarding the parameter values in Table~\ref{parameter_values} follow.
{\footnotesize
\begin{enumerate}
    \item\label{transmission} 
    The parameter $\beta$ strongly depends on the population behaviour.
    We take the value of $\beta$ from  \cite{shen2020modelling}, where they calibrated an SEIR model with isolation and estimated the transmission rate $\beta$, before lockdown, to be $0.7676$ (with a 95$\%$ confidence interval $(0.7403 , 0.7949)$).
    \item\label{latent} 
    The mean duration of the latent period can be computed using the estimates for the incubation period (i.e. from exposure to symptom onset) and the time from infectiousness onset to symptom onset, so it is reasonable to take $\tau^{-1}$ between 2 and 4 days. More precisely, in \cite{liang2020impacts} they fitted an SEIQR model to the data from Wuhan and estimated a latent period of duration 2.92 days with a 95\% CI of $(1.09, 5
    .28)$.
    \item\label{asymp}  
     In~\cite{wiki:diamond} they show the testing results on Diamond Princess passengers, a cruise ship that was quarantined in February-March 2020, at the beginning of the epidemic. Almost all passengers and crew members were tested, resulting in 410 asymptomatic infections among 696 positive-tested persons, which yields an asymptotic rate of $0.589$.
     In \cite{lavezzo2020suppression}, they studied the infection in the municipality of Vo', Italy. They estimated a median of asymptomatic cases of 44.8\%. 
     with a 95\% CI of (26.5,64.3).
     Other estimates were given in \cite{Daym1375,mizumoto2020estimating,nishiura2020estimation}.
    \item\label{rk:infectiouswindow}
    In \cite{Zhou2020}, they measured time from infectiousness onset to appearance of symptoms. It resulted in approximate 1 day for fever and 1-3 days for cough.
    Furthermore, it has been observed in clinical cases studied in~\cite{Woelfel_etal} that the contagious period may  start before the appearance of symptoms, and outlast the symptoms end. 
    \item\label{incubationWuhan} The reference \cite{backer2020incubation} estimates $\theta$ to be 6.4 based on travellers returning from Wuhan. 
    {In \cite{lauer2020incubation} it was estimated to be 5.1 days.
    Other estimates} were given in
    \cite{liang2020impacts}.
    \item\label{recovery1}
    The estimate of $\gamma_1$ is difficult, since for asymptomatic cases is hard to observe and track the time from exposure to recovery. \cite{Hu_etal_Asy} estimated 9.5 days for asymptomatic cases, while \cite{WHO-Ch} estimated 14 days for mild cases. So it is reasonable to assume $\gamma_1$ in the range 7.5 - 12, considering around 2 days between infectiousness onset and symptoms onset.
    \item\label{recovery2} 
    In \cite{Zhou2020} they measured viral shedding duration, and estimated a median of 20 days, with an {interquartile range} of $(17,24).$ Removing the approximately 2-day period from {infectiousness} onset to symptoms onset, we get {an IQR} for $\gamma_2^{-1}$ of $(15,22).$
    These values approximate the duration of  quarantine recommended to  positive-tested cases.
    \item\label{mu} 
    The rate $\mu$ depends on the percentage of infections that have been detected, since it is proportional to the ratio between confirmed cases and deaths.
    WHO Director-General's opening remarks at the media briefing of 3 March 2020~\cite{whoremarks3media2020} announced an estimated global death rate of 3.4\%. In some countries, like Italy, the ratio between deaths and confirmed cases up to May 2020 is larger that 0.1, while in others, like Israel, it is around 0.01. Regarding the time a person takes to die from COVID-19, in \cite{Zhou2020} they estimated $18.5$ days from infectiousness onset to death.
\item\label{pr} The values of $p$ and $r$ vary in each country/territory depending on the public policies and the population's compliance to these measures.  A detailed and real-time survey on the percentage of people under lockdown in each country can be found in~\cite{OxCGRT}.
\item\label{rho} As already mentioned in Remark \ref{RemarkTests}, $\rho$ represents the proportion of the infected asymptomatic population that is tested daily. In a realistic scenario, it would not be reasonable to set a too high value of $\rho$, let us say, over 0.5, because it would account for detecting  more than $50\%$ of the infected asymptomatic population daily.
\item\label{delta} It is not yet know ``at what point during the course of illness a test becomes positive'' (see \cite{HarvardMedicine}).  For the simulations we set $\delta$ to 1 and suppose that the tests detect the infection from  exposure.
\end{enumerate}
}

\subsection{Simulations for different scenarios}\label{subsec:Scenarios}


{In this subsection we consider several scenarios and show their outcomes. Many of the graphics are in logarithmic scale, given that the values represent portions of the population, and then can assume very small values.}

\subsubsection{Scenarios A}

{We consider the four different scenarios with the following characteristics:}

\vspace{5pt}

\noindent{\bf Scenario A$_1$:}
 no isolation, no testing among asymptomatic people
 
 \vspace{5pt}

\noindent{\bf Scenario A$_2$:} 
 20\%-isolation of 60\% of the population from day 31, no testing among asymptomatic people
 
 \vspace{5pt}

\noindent{\bf Scenario A$_3$:}
 20\%-isolation of 90\% of the population from day 31, no testing among asymptomatic people
 
 \vspace{5pt}

\noindent{\bf Scenario A$_4$:}
 20\%-isolation of 90\% of the population from day 31, intensive testing among asymptomatic people

\vspace{5pt}

\begin{figure}[h]
    \centering
    \begin{subfigure}[b]{.47\textwidth}
        \includegraphics[width=\textwidth]{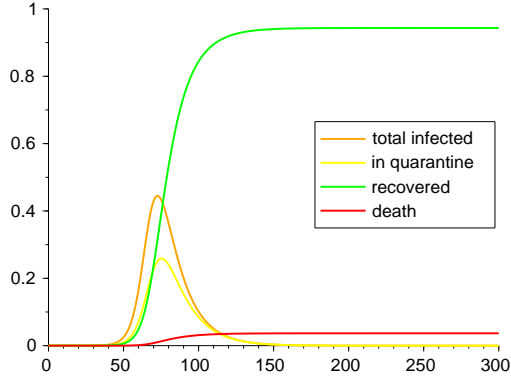}
        \caption{Scenario A$_1$}
        \label{A1}
    \end{subfigure}
    ~ 
    \begin{subfigure}[b]{.47\textwidth}
        \includegraphics[width=\textwidth]{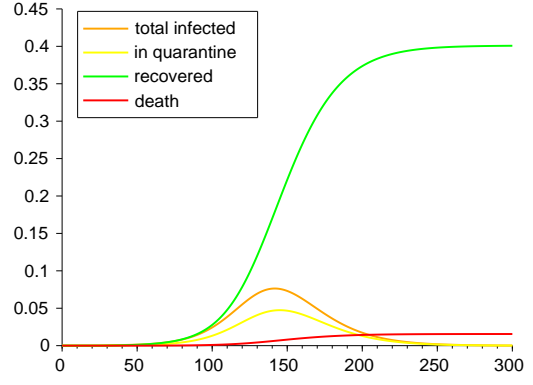}
        \caption{Scenario A$_2$}
        \label{A2}
    \end{subfigure}
    ~ 
    \begin{subfigure}[b]{.47\textwidth}
        \includegraphics[width=\textwidth]{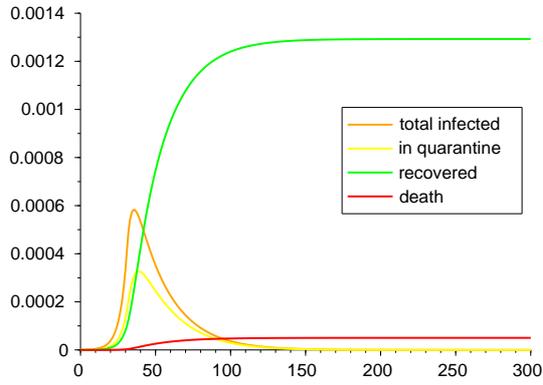}
        \caption{Scenario A$_3$}
        \label{A3}
    \end{subfigure}
    ~ 
    \begin{subfigure}[b]{.47\textwidth}
        \includegraphics[width=\textwidth]{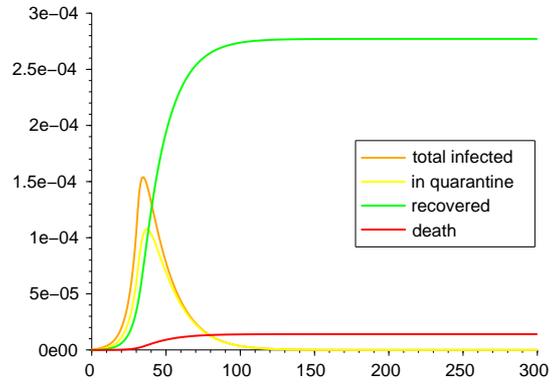}
        \caption{Scenario A$_4$}
        \label{A4}
    \end{subfigure}
    \caption{Scenarios A$_1$, A$_2$, $A_3$ and A$_4$.}\label{FigA14}
\end{figure}

\noindent
These Scenarios A can be seen as: no action, mild lockdown, strict lockdown and strict lockdown with testing among asymptomatic suspected cases.
As initial condition we take,  in all the scenarios,   one exposed case per million inhabitant, this is: 
\[
E_f(0)+E_r(0) = 1 \times 10^{-6},\quad 
S_f(0)+S_r(0) = 1-1 \times 10^{-6}.
\] 
 The remainder of the compartments start with value 0. Results and parameters for Scenario A are given in Table \ref{table:A14} and graphics in Figure \ref{FigA14}. We can observe the effect of the lockdown on the epidemic. The mild lockdown of A$_2$ reduces more than half of the infections w.r.t. the no action situation A$_1$, while  the strict lockdowns A$_3$ and A$_4$ induce a reduction of the order of $10^{-2}$ in  total recovered, deaths and positive tests. In particular, comparing A$_3$ and A$_4$ we can see that testing and consequent quarantine for positive-tested  asymptomatic cases not only reduces the infections and deaths more than 66\%, but also {the duration} of the epidemic.

\begin{remark}[About Scenario A$_2$]
United States had 15 confirmed infections by February 15th 2020~\cite{worldometers-us}. Many states started their lockdown between March 15th and March 20th, which put around 60\% of the population in isolation \cite{OxCGRT}. We can assume that day 0 is February 15th, then day 31 would be in the middle point of the interval when lockdowns started. Day 80 (May 6th) US had 72,287 confirmed deaths, while in Scenario A$_2$ gives  91,231, that is larger but not far. Scenario A$_2$ could be a good approximation of the situation in the US until the beginning of May 2020, and  the excess in the computed deaths in comparison to real data suggests that deaths could be underreported by $26\%$, which is coherent with some recent studies on underreported deaths (see {\em e.g.} \cite{newyork-missingdeaths}). 
\end{remark}

\begin{table}[H]
\begin{center}
\resizebox{0.75\textwidth}{!}{\begin{tabular}{|c|c|c|c|c|}
\hline
{\bf Par.} & {\bf A$_1$} & {\bf  A$_2$} & {\bf A$_3$} & {\bf  A$_4$}  \\[0.5ex]
\hline
$\beta$ & \multicolumn{4}{|c|}{0.7676} 
\\ \hline
$p$  & 0 & 
\begin{tabular}{c}
$0$ if $t\leq 31$\\  $0.6$ if $t>31$
\end{tabular}& 
\multicolumn{2}{|c|}{\begin{tabular}{c}
$0$ if $t\leq 31$\\  $0.9$ if $t>31$
\end{tabular}}
\\ \hline
$r$ & 1 
&  
\multicolumn{3}{|c|}{\begin{tabular}{c}
$1$ if $t\leq 31$\\  $0.2$ if $t>31$
\end{tabular}} 
\\ \hline
$\rho$ & \multicolumn{3}{|c|}{0} & 0.05 \\
\hline
$\delta$ &\multicolumn{4}{|c|}{1}\\
\hline
$\tau$ &\multicolumn{4}{|c|}{1/3.2} \\
\hline
$\sigma$ & \multicolumn{4}{|c|}{1/2} \\
\hline
$\theta$ & \multicolumn{4}{|c|}{1/5.2} \\
\hline
$\gamma_1$ & \multicolumn{4}{|c|}{1/8} \\
\hline
$\gamma_2$ & \multicolumn{4}{|c|}{1/16} \\
\hline
$\mu$ & \multicolumn{4}{|c|}{0.058/14} \\
\hline
$\alpha$ & \multicolumn{4}{|c|}{0.4} \\
\hline 
$\mathcal{R}_0$ & 2.51 & 
\begin{tabular}{c}
$2.51$ if $t\leq 31$\\  $1.4$ if $t>31$
\end{tabular}
&
\begin{tabular}{c}
$2.51$ if $t\leq 31$\\  $0.69$ if $t>31$
\end{tabular}& 
\begin{tabular}{c}
$1.92$ if $t\leq 31$\\ 
$0.52$ of $t>31$
\end{tabular}\\
\hline
\begin{tabular}{c}
   peak day  \\
   (for $Q$)
\end{tabular} & 76 & 146 &  40 & 38  \\
\hline
recovered & $9.43 \times 10^{-1}$ 
& $4.01 \times 10^{-1}$
& $1.29 \times 10^{-3}$ & 
$2.77 \times 10^{-4}$ \\
\hline
deaths & $3.66 \times 10^{-2} $ 
& $1.56 \times 10^{-2} $
& $5.02 \times 10^{-5}$ &$1.41 \times 10^{-5}$ \\
\hline
positive tests & $5.88 \times 10^{-1}$
& $2.5 \times 10^{-1}$
& $8.06 \times 10^{-4}$ &  $2.26 \times 10^{-4}$ \\
\hline
\begin{tabular}{c}
   ending day  \\
     ($Q\leq 10^{-9}$)
\end{tabular} & 377 & $>$500 & 314 & 225\\
\hline
\end{tabular}}\\
\caption{Scenarios A$_1$, A$_2$, $A_3$ and A$_4$. Parameters and epidemics output}
\label{table:A14}
\end{center}
\end{table}

For Scenario A$_4$ we make a comparison of infections for the two groups: the {active one (that continues with the usual contact rate)} and the one in $r$-isolation. By comparing the infections' curves and the cumulative infections, we can give an estimate on the lower chance that people in $r$-isolation have to get exposed. In this particular scenario, people that are not in isolation have nearly 6 times more chance to be infected. See the graphs in Figure \ref{sec:R0}, were we show the curves of infections and cumulative infections for each group, normalized by the proportions $1-p$, $p.$

\begin{figure}[H]
    \centering
    \begin{subfigure}[b]{.65\textwidth}
        \includegraphics[width=\textwidth]{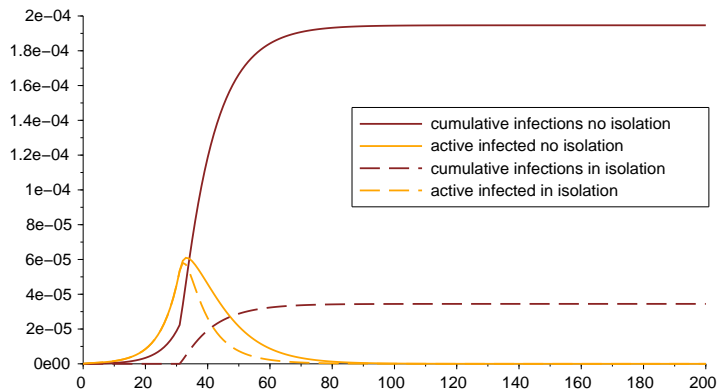}
    \end{subfigure}
    ~ 
    \caption{Comparison of infections for population in and out isolation}\label{A4comparison}
\end{figure}

\subsubsection{Scenarios B: different restriction level of lockdown}

We next consider the following four scenarios in which we vary the values of the portion $p$ of people under lockdown and their level $r$ of restriction of social contacts. 

\vspace{5pt}

 \noindent{\bf Scenario B$_1$:}  
 50\%-isolation of 50\% of the population from day 35  

 \vspace{5pt}

 \noindent{\bf Scenario B$_2$:} 
 40\%-isolation of 65\% of the population from day 35  
 
 \vspace{5pt}

 \noindent{\bf Scenario B$_3$:}  
 30\%-isolation of 80\% of the population from day 35  
 
 \vspace{5pt}

 \noindent{\bf Scenario B$_4$:}  
 20\%-isolation of 90\% of the population from day 35  
 
 \vspace{5pt}

\begin{figure}[H]
    \centering
    \begin{subfigure}[b]{.49\textwidth}
        \includegraphics[width=\textwidth]{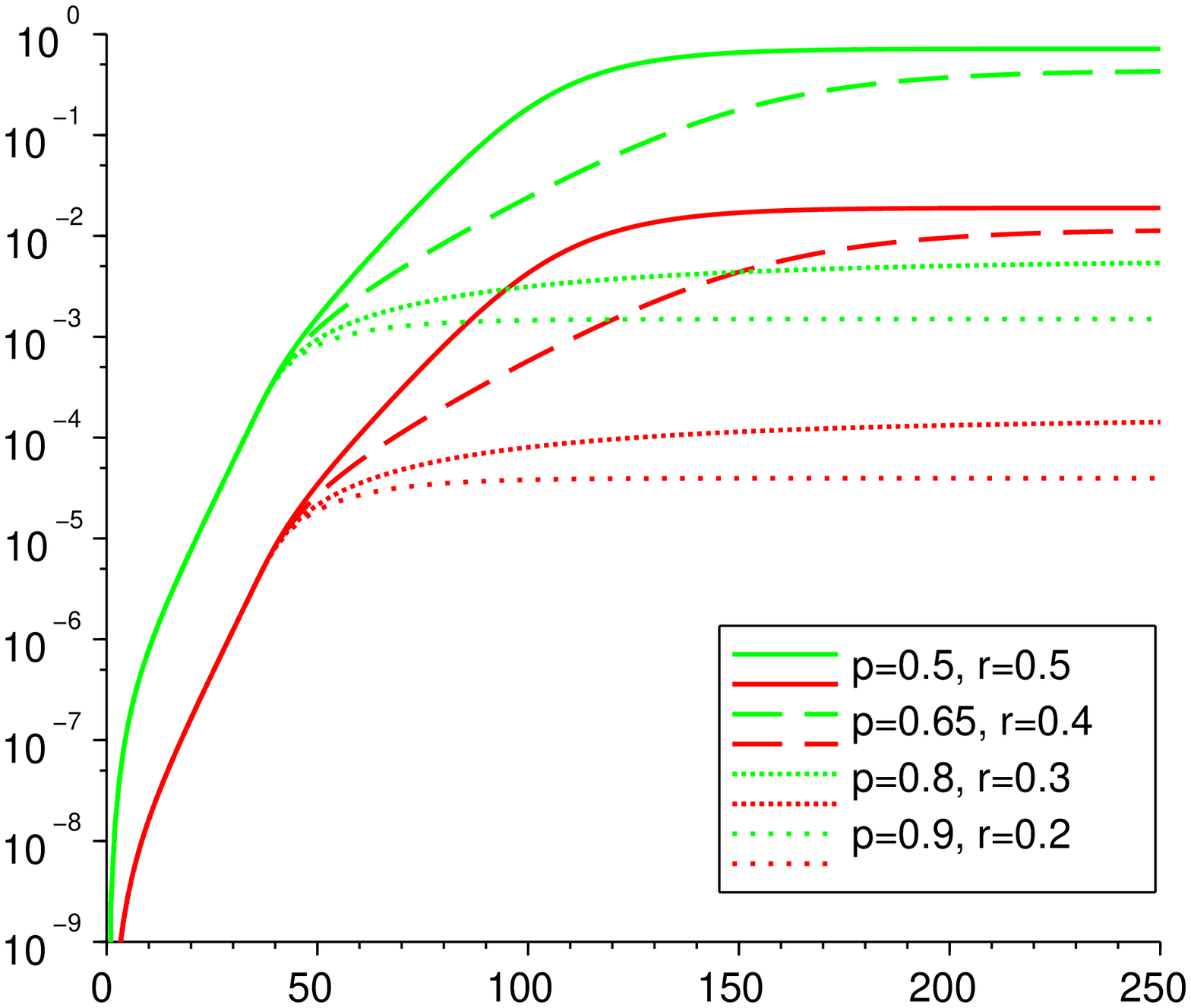}
        \caption{{\color{green}Recovered in  green}, {\color{red} dead in red}}
        \label{BRD}
    \end{subfigure}
    ~ 
    \begin{subfigure}[b]{.47\textwidth}
        \includegraphics[width=\textwidth]{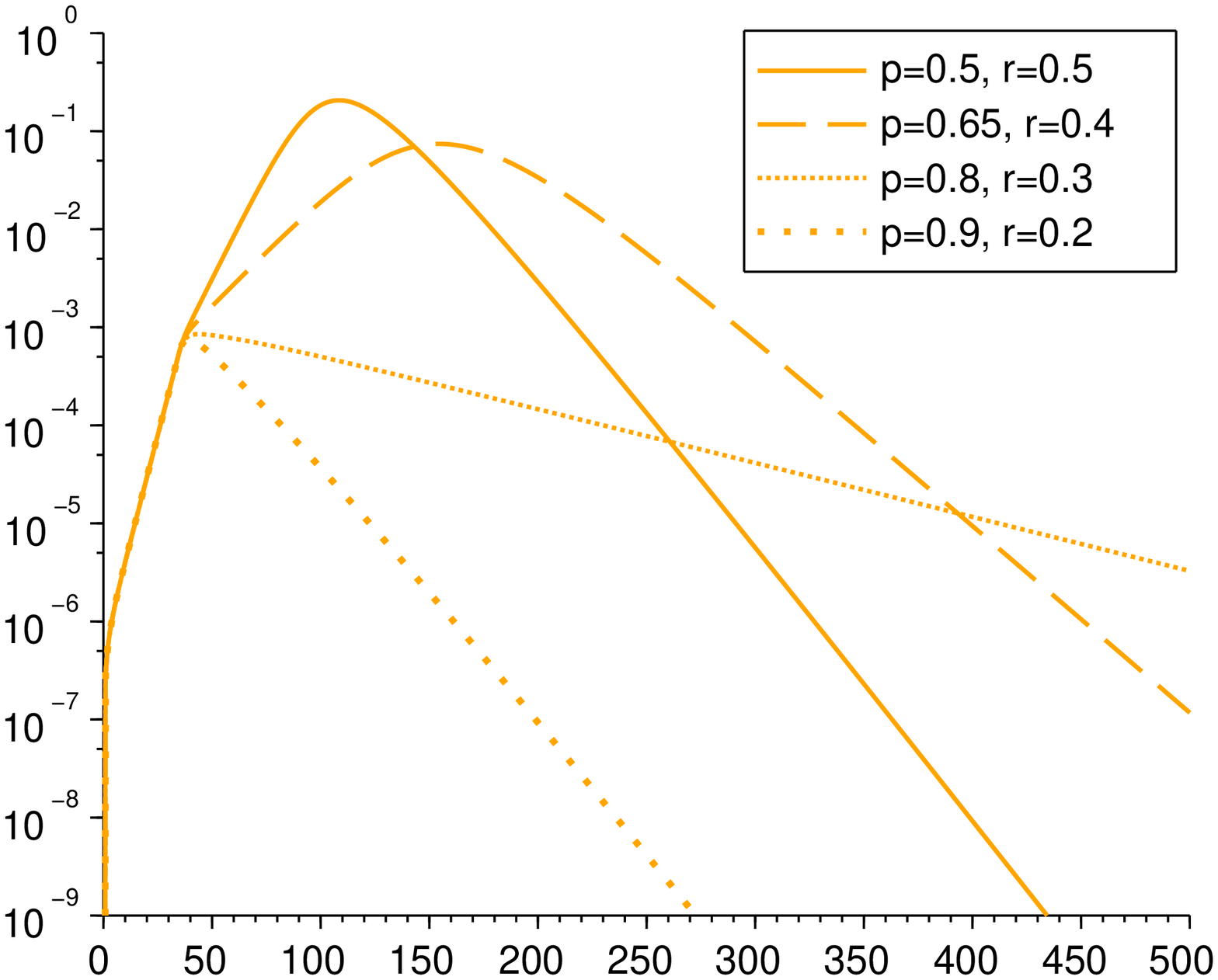}
        \caption{Infected}
        \label{BI}
    \end{subfigure}
     \caption{Scenarios B$_1$, B$_2$, B$_3$ and B$_4$}\label{FigB}
\end{figure}

 \noindent
We measure the outcomes. The parameters and results are given in Table \ref{scenariosB13}, and graphics in Figure \ref{FigB}. The parameters that are not specified in Table \ref{scenariosB13}, are repeated from Table \ref{table:A14}.
For these four scenarios we consider the same testing rate $\rho$ among asymptomatic cases.

Scenarios B$_1$ and B$_2$ show cases in which the restrictions are not strong enough. Indeed, in both cases the basic reproduction number $\mathcal{R}_0$ remains above 1 also after the lockdown intervention (see Table~\ref{scenariosB13}), and the infection reaches 71.6\% and 43.6\% of the population, causing the death of 1.89\% and 1.15\% of the population, respectively, which is a catastrophic outcome. Comparing these four scenarios, we shall deduce that, in order to be effective in containing the outbreak, the lockdown shall address at least $80\%$ of the population reducing their contact rate to about~$30\%$ of their usual contacts. Indeed, in the scenario B$_3$, the basic reproduction number becomes 0.93 after day 35, meaning that loosening the restrictions of this scenario (while keeping all other parameters unchanged) might turn the $\mathcal{R}_0$ above 1.

\begin{table}
\begin{center}
\resizebox{0.75\textwidth}{!}{\begin{tabular}{|c|c|c|c|c|}
\hline
{\bf Par.} & {\bf  B$_1$} & {\bf  B$_2$} & {\bf  B$_3$} & {\bf  B$_4$}\\[0.5ex]
\hline
$\beta$ & \multicolumn{4}{|c|}{0.7676} \\
\hline
$\mu$ & \multicolumn{4}{|c|}{0.034/14} \\
\hline
$p$  &  
\begin{tabular}{c}
$0$ if $t\leq 35$\\  $0.5$ if $t>35$
\end{tabular}
& 
\begin{tabular}{c}
$0$ if $t\leq 35$\\  $0.65$ if $t>35$
\end{tabular}  
&
\begin{tabular}{c}
$0$ if $t\leq 35$\\  $0.8$ if $t>35$
\end{tabular}
&
\begin{tabular}{c}
$0$ if $t\leq 35$\\  $0.9$ if $t>35$
\end{tabular}\\
\hline
$r$ &  
\begin{tabular}{c}
$1$ if $t\leq 35$\\  $0.5$ if $t>35$
\end{tabular}
& 
\begin{tabular}{c}
$1$ if $t\leq 35$\\  $0.4$ if $t>35$
\end{tabular}  
&
\begin{tabular}{c}
$1$ if $t\leq 35$\\  $0.3$ if $t>35$
\end{tabular}
&
\begin{tabular}{c}
$1$ if $t\leq 35$\\  $0.2$ if $t>35$
\end{tabular}\\
\hline
$\rho$ & \multicolumn{4}{|c|}{0.02} \\
\hline
$\mathcal{R}_0$ & 
\begin{tabular}{c}
$2.24$ if $t\leq 35$\\  $1.61$ if $t>35$
\end{tabular} & \begin{tabular}{c}
$2.24$ if $t\leq 35$\\  $1.23$ if $t>35$
\end{tabular}& 
\begin{tabular}{c}
$2.24$ if $t\leq 35$\\ 
$0.93$ of $t>35$
\end{tabular}&
\begin{tabular}{c}
$2.24$ if $t\leq 35$\\ 
$0.61$ of $t>35$
\end{tabular}\\
\hline
\begin{tabular}{c}
   peak day  
\end{tabular} & 113 & 159 & 51 & 43 \\
\hline
recovered & $7.16 \times 10^{-1}$ & $4.36 \times 10^{-1}$ & $5.78 \times 10^{-3}$ & $1.45 \times 10^{-3}$ \\
\hline
deaths & 
$1.89 \times 10^{-2} $ & $1.15\times 10^{-2}$ &
$1.53  \times 10^{-4}$ &
$3.96 \times 10^{-5}$\\
\hline
positive tests & 
$5.06 \times 10^{-1}$ & 
$3.08 \times 10^{-1}$ &  $4.09 \times 10^{-3}$ &
$1.06 \times 10^{-3}$\\
\hline
\begin{tabular}{c}
   ending day  
\end{tabular} 
& 436 &  $>500$
&  $>500$  & 274\\
\hline
\end{tabular}}\\
\caption{Scenarios B$_1$, B$_2$, B$_3$ and B$_4$. Parameters and epidemic outputs}
\label{scenariosB13}
\end{center}
\end{table}

\subsubsection{Scenarios C: early vs. late lockdown}

We now compare two situations, one in which lockdown starts  immediately, just 21 days after the first confirmed cases, and the other for which lockdown  starts four weeks later. More precisely, we consider the following two scenarios and measure the different outputs:

\vspace{5pt}

\noindent{\bf Scenario C$_1$:} 
20\%-isolation of 90\% of the population from day 21

\vspace{5pt}

\noindent{\bf Scenario C$_2$:}
 20\%-isolation of 90\% of the population from day 49
 
 \vspace{5pt}

\begin{figure}[h]
    \centering
    \begin{subfigure}[b]{.47\textwidth}
        \includegraphics[width=\textwidth]{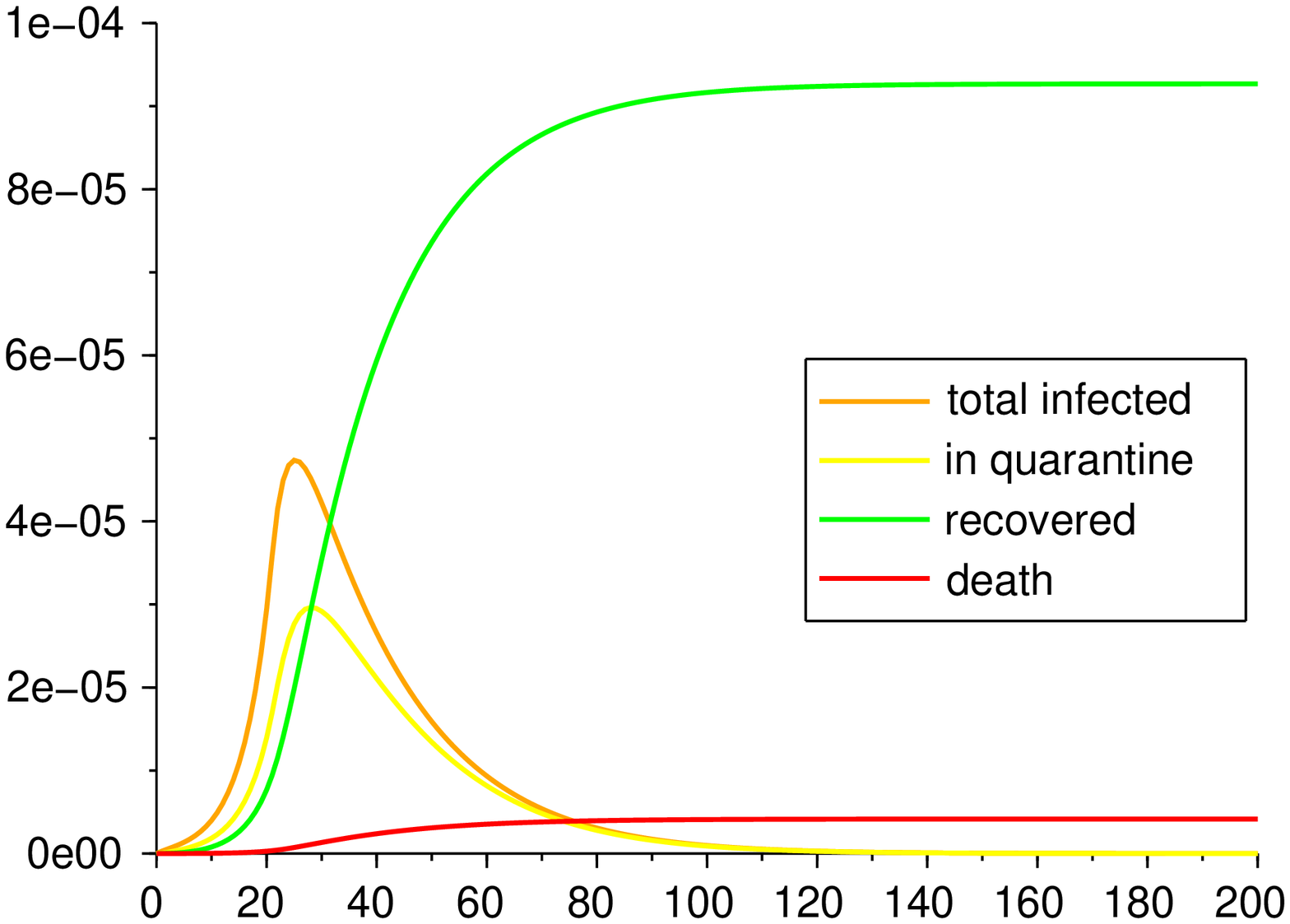}
        \caption{Scenario C$_1$}
        \label{C1}
    \end{subfigure}
    ~ 
    \begin{subfigure}[b]{.47\textwidth}
        \includegraphics[width=\textwidth]{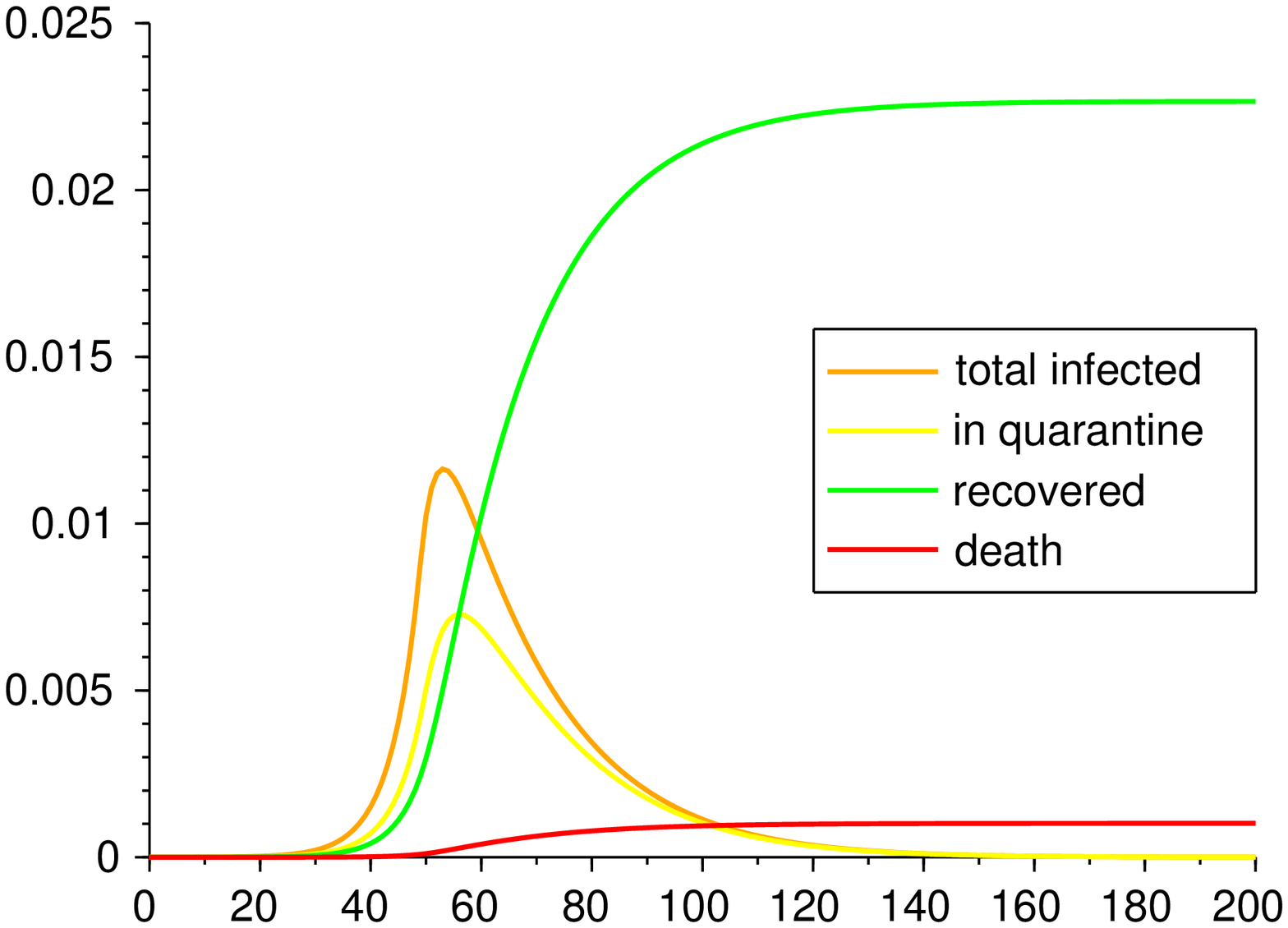}
        \caption{Scenario C$_2$}
        \label{fig:tiger}
    \end{subfigure}
    \caption{Scenarios C$_1$ and C$_2$.}\label{FigC12}
\end{figure}

\noindent The parameters and outputs are given in Table \ref{scenariosC12} and Figure \ref{FigC12}. It is evident the impact of delaying the beginning of lockdown on the final outcome: the numbers of recovered and deaths in the Scenario C$_2$ are of the order of $10^3$ {times} those of the Scenario C$_1$. As an example, from Table \ref{scenariosC12} one notice that, at the end of the epidemic, the Scenario C$_1$ counts {4.2} deaths per million inhabitants, while the Scenario C$_2$ faces 1020 deaths per million. Moreover, the epidemic in Scenario $C_2$ lasts about 110 days more than in Scenario $C_1$, thus also undergoes worst economic consequences of the lockdown.

\begin{table}
\begin{center}
\scalebox{0.9}{\begin{tabular}{|c|c|c|}
\hline
{\bf Par.} & {\bf C$_1$} & {\bf  C$_2$}\\[0.5ex]
\hline
$\beta$ & \multicolumn{2}{|c|}{0.7676} \\
\hline
$\mu$ & \multicolumn{2}{|c|}{0.058/14} \\
\hline
$p$  &  
\begin{tabular}{c}
$0$ if $t\leq 21$\\  $0.9$ if $t>21$
\end{tabular}
&
\begin{tabular}{c}
$0$ if $t\leq 49$\\  $0.9$ if $t>49$
\end{tabular}
\\
\hline
$r$ &  
\begin{tabular}{c}
$1$ if $t\leq 21$\\  $0.2$ if $t>21$
\end{tabular}
&
\begin{tabular}{c}
$1$ if $t\leq 49$\\  $0.2$ if $t>49$
\end{tabular}\\
\hline
$\rho$ &
\multicolumn{2}{|c|}{0.02}\\
\hline
$\mathcal{R}_0$ & 
\begin{tabular}{c}
$2.24$ if $t\leq 21$\\  $0.61$ if $t>21$
\end{tabular} & 
\begin{tabular}{c}
$2.24$ if $t\leq 49$\\  $0.61$ if $t>50$
\end{tabular}\\
\hline
\begin{tabular}{c}
   peak day  
\end{tabular} & 29 & 57 \\
\hline
recovered & $9.17 \times 10^{-5}$ & $2.27 \times 10^{-2}$ \\
\hline
deaths & 
$4.2 \times 10^{-6} $ & $1.02\times 10^{-3}$\\
\hline
positive tests & 
$6.67 \times 10^{-5}$ & 
$1.63 \times 10^{-2}$\\
\hline
   ending day  
&  213 
& 323\\
\hline
\end{tabular}}\\
\caption{Scenarios C$_1$ and C$_2$: early lockdown vs. late lockdown. Parameters and epidemic outputs}
\label{scenariosC12}
\end{center}
\end{table}

\subsubsection{Scenarios D: early  testing vs. late  testing}

{We now want to asses the impact of testing timing. For this, we} consider the following two scenarios and measure the different outputs:\\

\vspace{5pt}

\noindent{\bf Scenario D$_1$:} 
20\%-isolation of 80\% of the population from day 50,
efficient testing before day 50, reduced testing after

\vspace{5pt}

\noindent{\bf Scenario D$_2$:} 
 20\%-isolation of 80\% of the population from day 50,
 few testing before day 50, massive testing after 

\vspace{5pt}

\begin{figure}[H]
    \centering
    \begin{subfigure}[b]{.47\textwidth}
        \includegraphics[width=\textwidth]{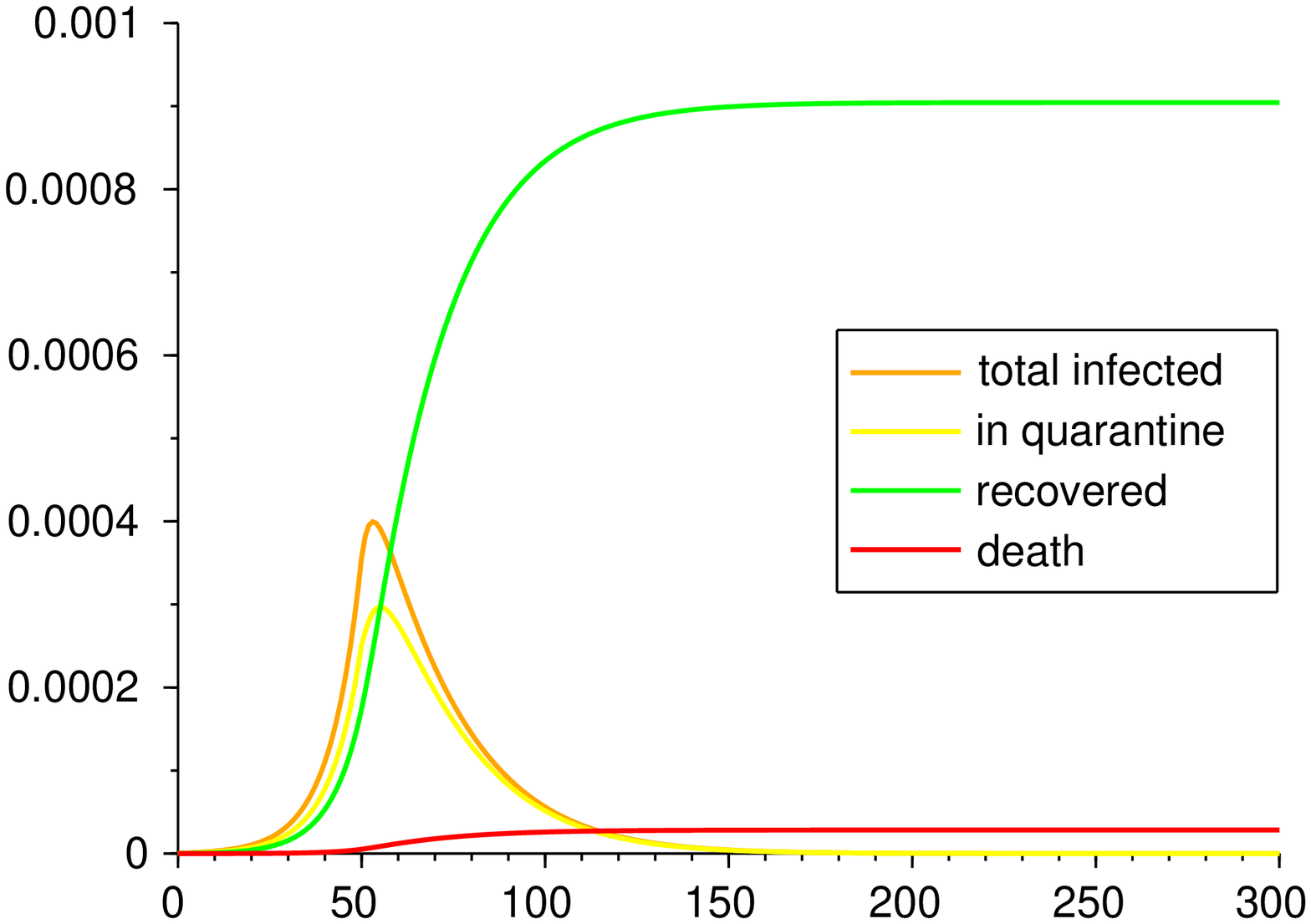}
        \caption{Scenario D$_1$}
        \label{C1}
    \end{subfigure}
    ~ 
    \begin{subfigure}[b]{.47\textwidth}
        \includegraphics[width=\textwidth]{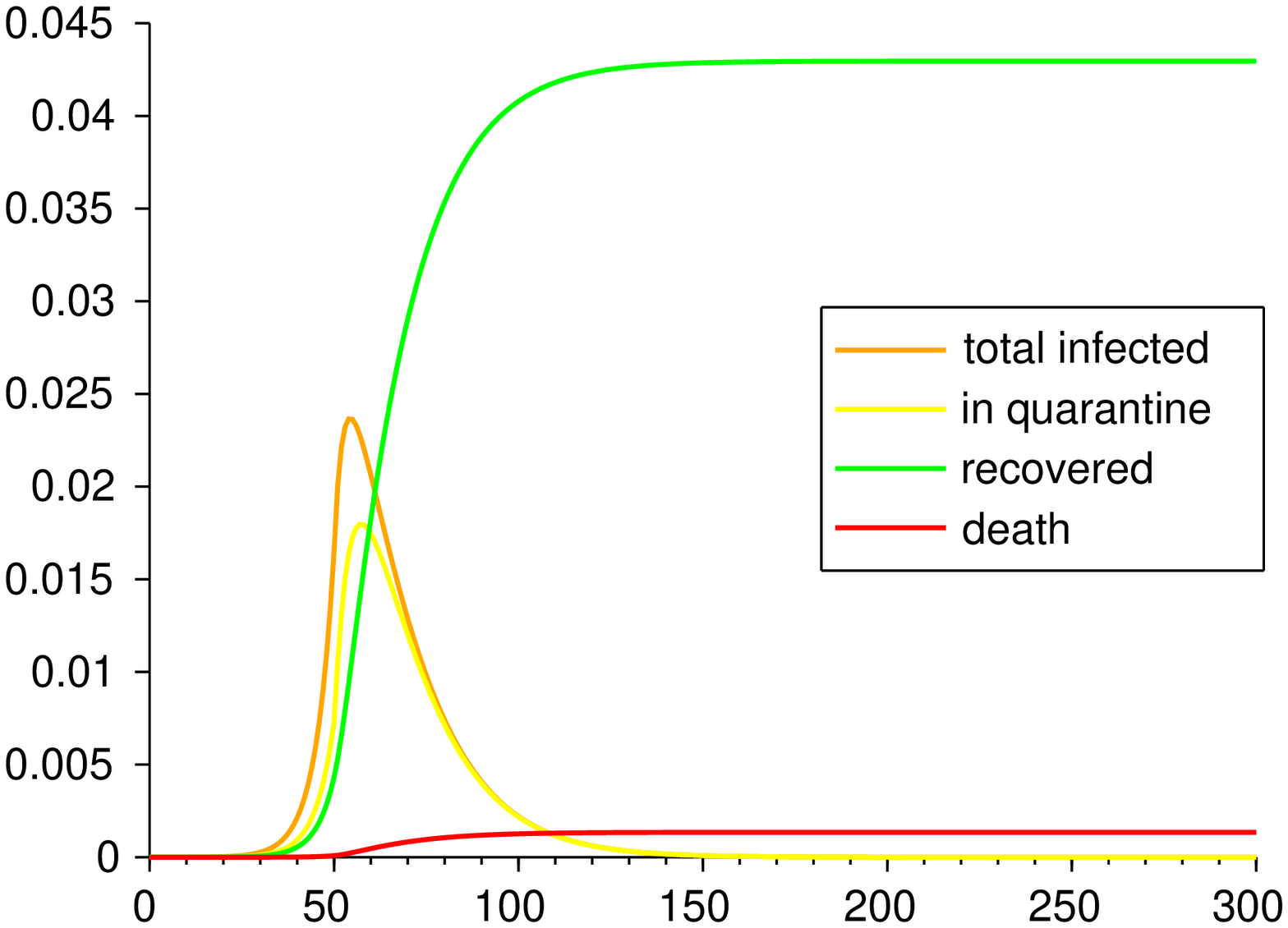}
        \caption{Scenario D$_2$}
        \label{fig:tiger}
    \end{subfigure}
    \caption{Scenarios D$_1$ and D$_2$.}\label{FigD12}
\end{figure}

\noindent The parameter values and outcomes of the epidemic in Scenarios D$_1$ and D$_2$ are given in Table~\ref{scenariosD12} and figures in Figure \ref{FigD12}. It can be seen the cost in infection and lives it has to start testing late. It is worth noticing that, in spite of a higher total number of tests carried out in the Scenario D$_2$, the strategy adopted in the Scenario D$_1$ attains a considerably better outcome: indeed, the infections and deaths of Scenario D$_2$ are of the order of 10$^2$ w.r.t. the ones in Scenario D$_1,$ and the only difference was doing efficient testing at the beginning of the epidemic.

 \begin{table}[h]
\begin{center}
\scalebox{0.9}{\begin{tabular}{|c|c|c|}
\hline
{\bf Par.} & {\bf D$_1$} & {\bf D$_2$}\\[0.5ex]
\hline
$\beta$ & \multicolumn{2}{|c|}{0.7676} \\
\hline
$\mu$ & \multicolumn{2}{|c|}{0.034/14} \\
\hline
$p$  &  
\multicolumn{2}{|c|}{\begin{tabular}{c}
$0$ if $t\leq 50$\\  $0.8$ if $t>50$
\end{tabular}}
\\
\hline
$r$ &  
\multicolumn{2}{|c|}{\begin{tabular}{c}
$1$ if $t\leq 50$\\  $0.2$ if $t>50$
\end{tabular}}\\
\hline
$\rho$ &
\begin{tabular}{c}
$0.1$ if $t\leq 50$\\  $0.05$ if $t>50$
\end{tabular}
&
\begin{tabular}{c}
$0.01$ if $t\leq 50$\\  $0.1$ if $t>50$
\end{tabular}\\
\hline
$\mathcal{R}_0$ & 
\begin{tabular}{c}
$1.53$ if $t\leq 50$\\  $0.9$ if $t>50$
\end{tabular} & \begin{tabular}{c}
$2.37$ if $t\leq 50$\\  $0.57$ if $t>50$
\end{tabular}\\
\hline
\begin{tabular}{c}
   peak day  
\end{tabular} & 56 & 58\\
\hline
recovered & $9.04 \times 10^{-4}$ & $4.29 \times 10^{-2}$ \\
\hline
deaths & 
$2.83 \times 10^{-5} $ & $1.35\times 10^{-3}$\\
\hline
positive tests & 
$7.58 \times 10^{-4}$ & 
$3.6 \times 10^{-2}$\\
\hline
   ending day  
&   $302$ 
& 327\\
\hline
\end{tabular}}\\
\caption{Scenarios D$_1$ and D$_2$: early efficient testing vs. late massive testing. Parameters and epidemic outputs}
\label{scenariosD12}
\end{center}
\end{table}

\subsection{Scenarios E: different testing rates}

\begin{figure}[H]
    \centering
    \begin{subfigure}[b]{.47\textwidth}
        \includegraphics[width=\textwidth]{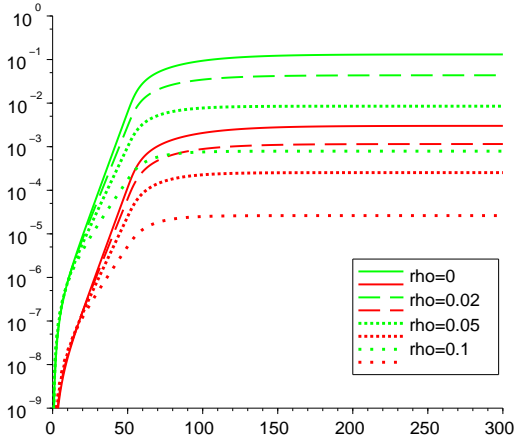}
        \caption{{\color{green}Recovered in  green}, {\color{red} dead in red}}
        \label{ERD}
    \end{subfigure}
    ~ 
    \begin{subfigure}[b]{.49\textwidth}
        \includegraphics[width=\textwidth]{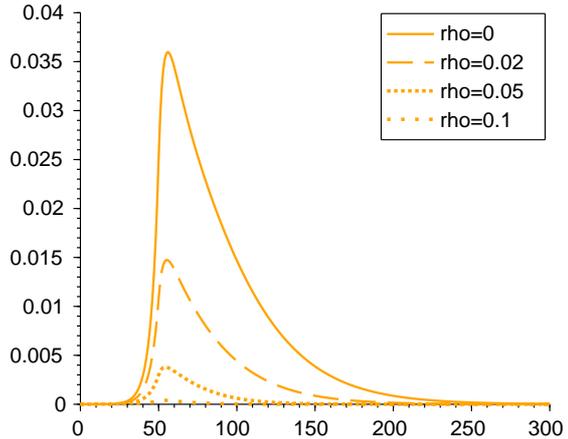}
        \caption{Infected}
        \label{EI}
    \end{subfigure}
     \caption{Scenarios E$_1$, E$_2$, E$_3$ and E$_4$}\label{FigE}
\end{figure}

Now we fix the parameters $\beta, \mu, p, r$ as in Table \ref{scenariosD12} and we vary only $\rho$ to take the four different values 0, 0.02, 0.05 and 0.1 over the whole time period. We get the outcome of Figure \ref{FigE}.
 From the comparison among these four scenarios, we realize that a high value of $\rho$, as the result of an efficient  tracing and testing strategy, may reduce the number of cumulative infected individuals and deaths of an order {of} $10^2$.

The simulations were done with Python and all the codes are in the GitHub repository\\ \href{https://github.com/lucasmoschen/covid-19-model}{github.com/lucasmoschen/covid-19-model}.

\section{Conclusions}\label{sec:conclusions}

In the paper we present an SEIR model with Asymptomatic and Quarantined compartments to describe the recent and ongoing COVID-19 outbreak. Our model is intended to highlight the strength of three different non-pharmaceutical interventions imposed by public policies in containing the outbreak and the total number of disease-induced infections and deaths:
\begin{itemize}
    \item[-] reduction of contact rate for a given portion of the population;
    \item[-] enforced quarantine for confirmed infectious individuals;
    \item[-] testing among the population to detect also asymptomatic infectious individuals.
\end{itemize}

On one hand we show that, as expected, each of these interventions has a beneficial impact on flattening the curve of the outbreak. On the other hand, the comparison among different scenarios shows the remarkable efficacy of an early massive testing approach, when the limited number of infected individuals makes easier and more effective the tracing of recent contacts of the individual, as in Scenario D$_1$, and of a timely lockdown, although in the presence of few confirmed infected cases, as in Scenario C$_1$. In both situations, the timing of the intervention plays a crucial role on the incisiveness of the public safety policy.

In addition, we give an explicit representation of the basic reproduction number in terms of the several parameters of the model, which allows to describe its dependence on the features of the virus and on the implemented non-pharmaceutical interventions.

\begin{figure}
    \centering
    \begin{subfigure}[b]{.47\textwidth}
        \includegraphics[width=\textwidth]{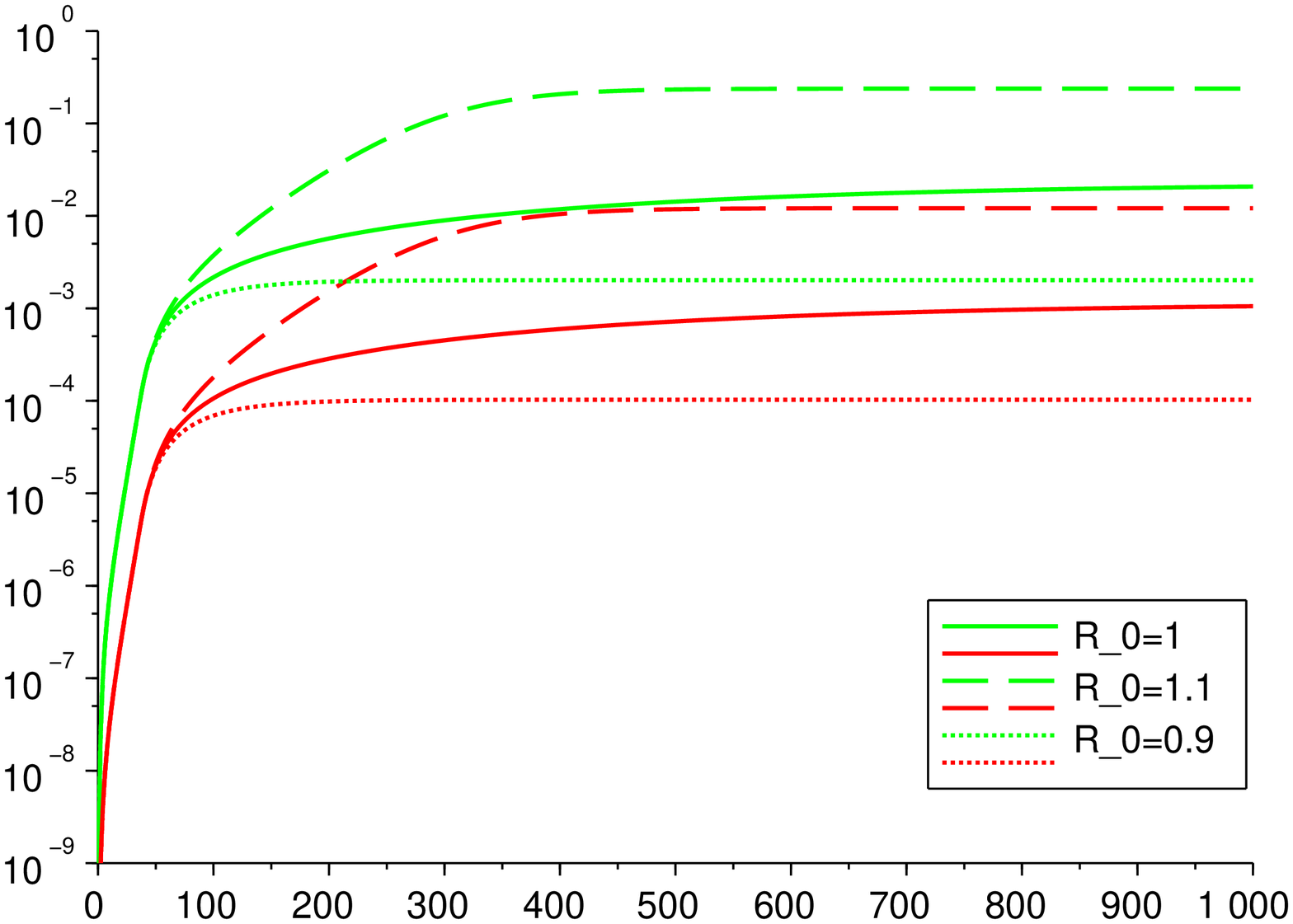}
        \caption{{\color{green}Recovered in  green}, {\color{red} dead in red}}
        \label{fig:recovereddeadR0}
    \end{subfigure}
    ~ 
    \begin{subfigure}[b]{.47\textwidth}
        \includegraphics[width=\textwidth]{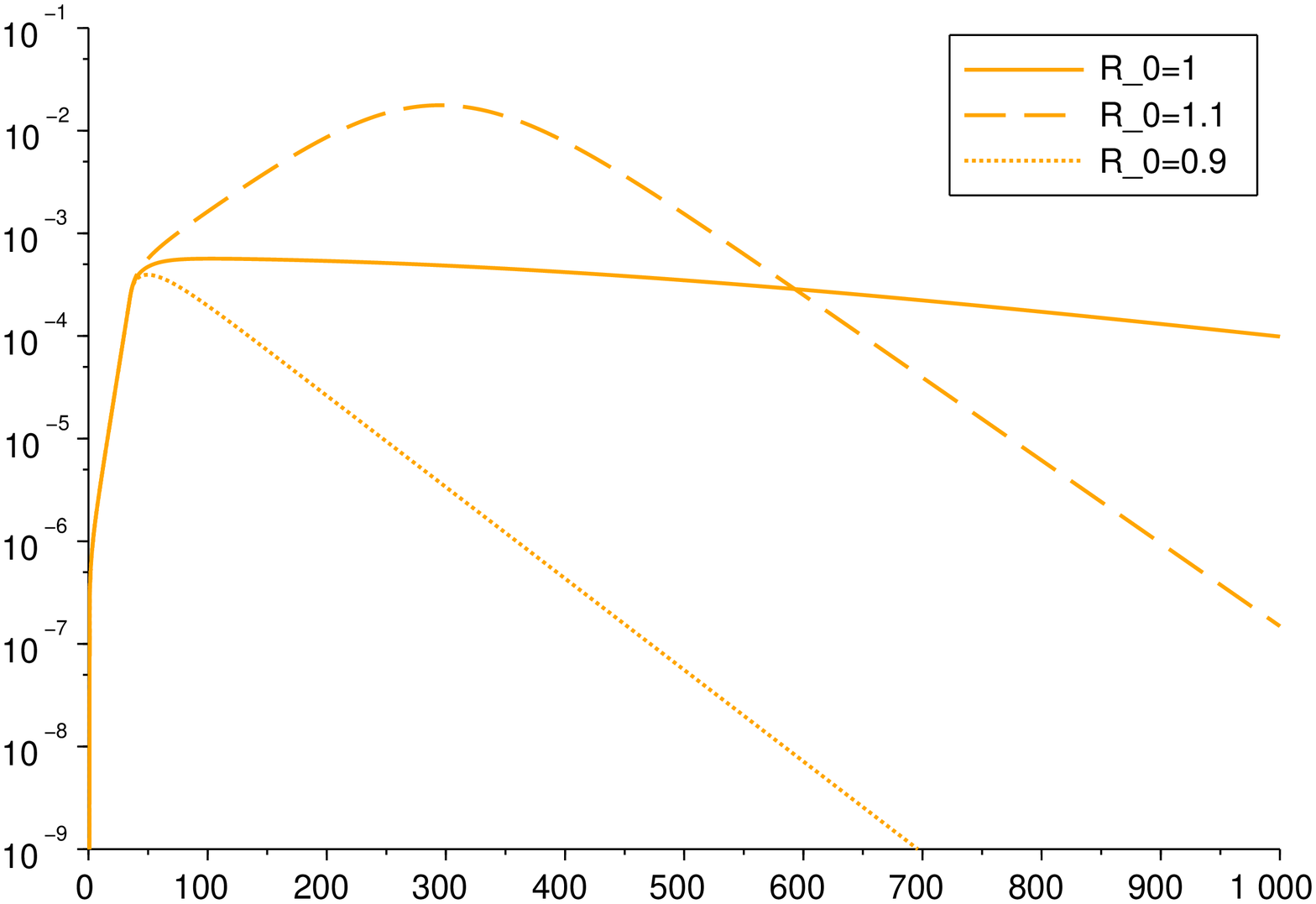}
        \caption{Infected}
        \label{fig:infectedR0}
    \end{subfigure}
    \caption{The impact of small variations on $\mathcal{R}_0$}\label{FigR0var}
\end{figure}

This description makes available a valuable tool to tune the public policies in order to control the outbreak of the epidemic, forcing $\mathcal{R}_0$ below the threshold $1$. However, considering the major effects of an enduring lockdown on the economy of the country that applies it, it is desirable to loosen the lockdown measures after the containment of the outbreak. Nevertheless, the decision makers and each individual shall be aware that a value of $\mathcal{R}_0$ only barely greater than $1$ would lead to an increase in the number of infected and dead by an order $2$ of magnitude, thus provoking the collapse of the relative national health system.
This is better explained by the following scenarios: consider a situation with constant testing $\rho = 0.05$ and no lockdown where, after the first 35 days of outbreak with a high $\mathcal{R}_0$ ($\approx 2$), the population gains awareness of the risk and manage to reduce its contact rate so as to steer $\mathcal{R}_0$ to either $0.9$, $1$ or $1.1$. 
Figure~\ref{FigR0var} illustrates the large deviations among the outcome of these three different situations.


In order to allow the population to circulate with no restrictions, 
it is necessary that herd immunity (see Remark \ref{rem:herd}) is achieved. The value that matters to compute this threshold of immunization is the basic reproduction number under no social distancing, which has been estimated in this article and in many others as being, in general, greater than 2.5. So achieving herd immunity  would imply to infect at least $60\%$  of the population, which would lead, with the current mortality rates, to 1-5\% of the population dying, which is, obviously, a catastrophic unwanted situation.
Hence, reinforcing what was said in the above paragraph, until a vaccine or treatment is not found, it is necessary to maintain the value of $\mathcal{R}_0$ below 1. Otherwise, the curve of infections will always be increasing.



\if{
\begin{table}
\begin{center}
\begin{tabular}{|c|c|c|c|}
\hline
{\bf Par.} & {\bf $\mathcal{R}_0$=1} & {\bf $\mathcal{R}_0$ =1.1} & {\bf $\mathcal{R}_0$ =0.9}\\[0.5ex]
\hline
$\beta$ 
& 
\begin{tabular}{c}
$0.7676$ if $t\leq 35$\\  $\hat\beta:=$ 0.2977333 if $t>35$
\end{tabular}
& 
\begin{tabular}{c}
$0.7676$ if $t\leq 35$\\  $1.15 \hat\beta$  if $t>35$
\end{tabular}
&
\begin{tabular}{c}
$0.7676$ if $t\leq 35$\\  $0.85 \hat\beta$  if $t>35$
\end{tabular} \\
\hline
$\mu$ & \multicolumn{3}{|c|}{0.058/14} \\
\hline
$p$  &  
\multicolumn{3}{|c|}{0}
\\
\hline
$r$ &  
\multicolumn{3}{|c|}{1}\\
\hline
$\rho$ &
\multicolumn{3}{|c|}{0.05}\\
\hline
\end{tabular}\\
\caption{Scenarios different $\mathcal{R}_0$}
\end{center}
\end{table}
}\fi

\appendix
\section{Appendix}

\subsection{Computing $\mathcal{R}_0$}\label{App1}

Recalling the model~\eqref{eq:SEIRwQ}, we are able to give an analytic expression of the basic reproduction number $\mathcal{R}_0$ associated to the system.

In order to do so, we assume to fix a time interval $[t_0,t_1]$ such that the coefficients $\beta(t)$, $r(t)$ and $\rho(t)$ are constant over $[t_0,t_1]$. This is coherent with the setting of the Section~\ref{sec:NumSim}, where we assume such coefficients to be piecewise constant functions, sharing the same switching times. Thus, the following procedure allows to evaluate the value of $\mathcal{R}_0$ for each time interval between two consecutive switching times.

It is well known that the reproduction number $\mathcal{R}_0$ is the crucial parameter to establish whether Disease Free Equilibria (DFE) are stable or not \cite{diekmann1990definition,vdDW02}.
We denote by $\mathcal{X}_s$ the set of DFE, which is given by
\begin{equation*}
\mathcal{X}_s = \left\{X\in \mathcal{C}_1 : E_f = E_r = I_f = I_{r} = A_f = A_{r} = Q = 0\right\}\, .
\end{equation*}
We can recast system~\eqref{eq:SEIRwQ} in the compact form
\begin{equation}\label{eq:abstsys}
    \Dot{X}(t) = f(X(t))
\end{equation}
by introducing
\begin{equation*}
    f(X) =
    \left(
\begin{smallmatrix}
    \beta S_f \left[I_f + A_f + r(I_{r} + A_r)\right] -\rho \delta E_f- \tau E_f \\[0.5ex]
r \beta S_r \left[I_f + A_f + r(I_{r} + A_r)\right] -\rho\delta E_r - \tau E_r\\[0.5ex]
\tau E_f - \sigma I_f - \rho I_f \\[0.5ex]
 \tau E_r - \sigma I_{r} - \rho I_{r} \\[0.5ex]
\sigma\alpha I_f - \rho A_f - \gamma_1 A_f \\[0.5ex]
\sigma\alpha I_{r} - \rho A_{r} - \gamma_1 A_r \\[0.5ex]
\sigma (1-\alpha) [I_f + I_{r}] + \rho \big(\delta(E_f+E_r)+I_f + I_{r} + A_f + A_{r}\big) - \gamma_2 Q - \mu Q \\[0.5ex]
 -\beta S_f [I_f + A_f + r( I_{r} + A_r)] \\[0.5ex]
-r \beta  S_r [I_f + A_f + r( I_{r} + A_r)]
    \end{smallmatrix}
    \right)\, .
\end{equation*}
The stability of~\eqref{eq:abstsys} around a DFE $X^*$ is related to the spectral properties of the linearized system around $X^*$, whose dynamics is ruled by the Jacobian $Df = (\partial f_i/\partial x_j)_{i,j = 1,\dots,9}$ of $f$. However, the high dimensionality of $Df(X)$ makes {it} difficult to develop an analytical analysis of its spectrum and its stability properties. We will therefore follow a different approach, deducing the value of $\mathcal{R}_0$ from the result in~\cite{vdDW02}, which ensures that $\mathcal{R}_0$ {is} given by the formula $\mathcal{R}_0 = \rho(FV^{-1})$, where $\rho(A)$ denotes the spectral radius of the matrix $A$. A comment on the applicability of the results in \cite{vdDW02} is given in Remark \ref{RemarkvdD} below.

Since $X^*$ is a DFE, we may assume that $X^* = (0,0,0,0,0,0,0,1-p,p)$, for some $p\in [0,1]$ representing the portion of population that is initially in the compartment $S_r$, while the remaining $1-p$ fraction of the population  is in $S_f$. Thus, in our setting,  the matrices $F$ and $V$ related to the dynamics~\eqref{eq:abstsys} are given by
\begin{equation*}
F = 
\left(
\begin{matrix}
0 & 0 & \beta (1-p) & r \beta (1-p) & \beta (1-p) & r \beta (1-p) & 0 \\[0.3ex]
0 & 0 & r \beta  p & r ^2\beta p & r \beta  p & r ^2\beta p & 0 \\[0.3ex]
0 & 0 & 0 & 0 & 0 & 0 & 0\\[0.3ex]
0 & 0 & 0 & 0 & 0 & 0 & 0\\[0.3ex]
0 & 0 & \sigma \alpha & 0 & 0 & 0 & 0\\[0.3ex]
0 & 0 & 0 & \sigma \alpha & 0 & 0 & 0\\[0.3ex]
0 & 0 & \sigma (1- \alpha) + \rho & \sigma (1- \alpha) + \rho & \rho & \rho & 0
\end{matrix}
\right)\, ,
\end{equation*}
\begin{equation*}
V = 
\left(
\begin{matrix}
\rho\delta + \tau & 0 & 0 & 0 & 0 & 0 & 0\\[0.3ex]
0 & \rho\delta + \tau & 0 & 0 & 0 & 0 & 0\\[0.3ex]
-\tau & 0 & \sigma + \rho & 0 & 0 & 0 & 0 \\[0.3ex]
0 & -\tau & 0 & \sigma + \rho & 0 & 0 & 0 \\[0.3ex]
0 & 0 & 0 & 0 & \rho + \gamma_1 & 0 & 0 \\[0.3ex]
0 & 0 & 0 & 0 & 0 & \rho + \gamma_1 & 0 \\[0.3ex]
-\rho\delta & -\rho\delta & 0 & 0 & 0 & 0  & \gamma_2 + \mu
\end{matrix}
\right)\, .
\end{equation*}
Since $V$ is non-singular, we compute
\begin{equation*}
V^{-1} = 
\left(
\begin{smallmatrix}
(\rho\delta + \tau)^{-1} & 0 & 0 & 0 & 0 & 0 & 0\\[0.3ex]
0 & (\rho\delta + \tau)^{-1} & 0 & 0 & 0 & 0 & 0\\[0.3ex]
\frac{\tau}{(\sigma + \rho)(\rho\delta + \tau)} & 0 & (\sigma + \rho)^{-1} & 0 & 0 & 0 & 0 \\[0.3ex]
0 & \frac{\tau}{(\sigma + \rho)(\rho\delta + \tau)} & 0 & (\sigma + \rho)^{-1} & 0 & 0 & 0 \\[0.3ex]
0 & 0 & 0 & 0 & (\rho + \gamma_1)^{-1} & 0 & 0 \\[0.3ex]
0 & 0 & 0 & 0 & 0 & (\rho + \gamma_1)^{-1} & 0 \\[0.3ex]
\frac{\rho\delta}{(\gamma_2 + \mu)(\rho\delta + \tau)} & \frac{\rho\delta}{(\gamma_2 + \mu)(\rho\delta + \tau)} & 0 & 0 & 0 & 0  & (\gamma_2 + \mu)^{-1}
\end{smallmatrix}
\right)\, .
\end{equation*}
Thus, one can easily compute the matrix $FV^{-1}$ and check that its characteristic polynomial is given by
\begin{equation*}
p(\lambda) = - \lambda^5 P_2(\lambda)\, ,
\end{equation*}
where $P_2(\lambda)$ is a second order polynomial of the form
\begin{equation*}
{P}_2(\lambda) = \lambda^2 - \frac{\beta \tau (1 - p + r^2 p)}{(\rho\delta + \tau)(\sigma + \rho)}\lambda - \frac{\sigma \alpha \tau \beta (1 - p + r^2 p)}{(\rho + \sigma)(\rho\delta + \tau)(\rho + \gamma_1)}\, .
\end{equation*}
${P}_2(\lambda)$ has one positive and one negative root, given by
\begin{equation*}
\lambda_{1/2} = \frac{1}{2}\left(
\frac{\beta \tau (1 - p + r^2 p)}{(\rho\delta + \tau)(\sigma + \rho)} \pm \sqrt{\Delta}
\right)\, ,
\end{equation*}
with
\begin{equation*}
\Delta = \left(\frac{\beta \tau (1 - p + r^2 p)}{(\rho\delta + \tau)(\sigma + \rho)}\right)^2 + \frac{4\sigma \alpha \tau \beta (1 - p + r^2 p)}{(\rho + \sigma)(\rho\delta + \tau)(\rho + \gamma_1)} > 0\, .
\end{equation*}
Since the term $\frac{\beta \tau (1 - p + r^2 p)}{(\rho\delta + \tau)(\sigma + \rho)}$ is positive, the value of $\mathcal{R}_0$ coincides with $\lambda_1$, i.e.,
\begin{equation}\label{eq:R0App}
\mathcal{R}_0 = \lambda_1 = \frac{1}{2}\left(
\frac{\beta \tau (1 - p + r^2 p)}{(\rho\delta + \tau)(\sigma + \rho)} + \sqrt{\Delta}
\right)\, .
\end{equation}
This is an analytic expression of $\mathcal{R}_0$, which shows its explicit dependence on the different parameters of model~\eqref{eq:SEIRwQ}.
Proposition~\ref{PropR0} gives a convenient equivalent condition to ensure the stability of DFE.

\begin{remark}
\label{RemarkvdD}
In order to directly apply the results in \cite{vdDW02}, it is required that the eigenvalues of $Df(X^*)$ have negative values and, under this assumption, the asymptotic stability of the DFE is established. In our case, the matrix $Df(X^*)$ has zero as an eigenvalue of double multiplicity, with associated eigenvectors in the directions of the last two variables, these being $S_f$ and $S_r.$ It is not hard to see that the results in \cite{vdDW02} hold for our system  by simply modifying asymptotic stability to stability in the directions of the susceptible compartments, which has no consequence in the meaning of the threshold $\mathcal{R}_0$.
Alternatively, a way to force the system to comply all the technical assumptions from \cite{vdDW02} is adding birth and natural mortality to our model, which has no relevant impact in the results we showed (since the natural daily birth/death rates are of the order of $10^{-5}$, hence negligible w.r.t. the other parameters).
\end{remark}

\subsection{Proof of Proposition~\ref{PropR0}}\label{App2}
We remind that Proposition~\ref{PropR0} claims the following: for
$$
\varphi = \frac{\beta \tau [1 - (1 - r^2) p]}{(\rho\delta + \tau)(\sigma + \rho)} \; \qquad\text{ and }\qquad
\mathcal{T}_0 = \varphi\left(
1 + \frac{\sigma\alpha}{\rho + \gamma_1}
\right)\, ,
$$
$\mathcal{R}_0$ is smaller than (respectively, equal to or greater than) 1 if and only if the same relation holds for $\mathcal{T}_0$. Indeed, by a straightforward computation we realize that
\begin{multline*}
    \mathcal{R}_0\le 1 \quad \Longleftrightarrow\quad 
    \varphi + \sqrt{\varphi^2 + \frac{4\sigma\alpha}{\rho + \gamma_1}\varphi}\le 2\\[0.3ex]
    \Longleftrightarrow\quad \left(0 < \right) \sqrt{\varphi^2 + \frac{4\sigma\alpha}{\rho + \gamma_1}\varphi}\le 2 -\varphi
    \quad \overset{\ast}{\Longleftrightarrow}\quad \varphi^2 + \frac{4\sigma\alpha}{\rho + \gamma_1}\varphi\le \left(2 -\varphi\right)^2\\[0.3ex]
    \Longleftrightarrow\quad \frac{4\sigma\alpha}{\rho + \gamma_1}\varphi\le 4 - 4\varphi
    \quad \Longleftrightarrow\quad
    \varphi\left(1 + \frac{\sigma\alpha}{\rho + \gamma_1}\right)\le 1
    \quad \Longleftrightarrow\quad
    \mathcal{T}_0 \le 1\; .
\end{multline*}
Observe that the implication $\ \Longleftarrow\ $ in the equivalence $\overset{\ast}{\ \Longleftrightarrow\ }$ holds true because
$$
\mathcal{T}_0\le 1\quad \Longrightarrow\quad \varphi \le \frac{\rho + \gamma_1}{\rho + \gamma_1 + \sigma\alpha} \le 1\, ,
$$
thus $|2 - \varphi| = 2 - \varphi$. In particular, the same chain of relations holds with the equal sign. Finally, since $\mathcal{R}_0\le 1$ is equivalent to $\mathcal{T}_0\le 1$, then also $\mathcal{R}_0 > 1$ is equivalent to $\mathcal{T}_0 > 1$.

In addition, let us notice that, if $\varphi > 1$, then both $\mathcal{R}_0 > 1$ and $\mathcal{T}_0 > 1$. Indeed, from the definition of $\mathcal{T}_0$, since $\frac{\sigma\alpha}{\rho + \gamma_1} \geq 0$, we have that $\mathcal{T}_0 \geq \varphi >1$, and thus also $\mathcal{R}_0 > 1$.

\subsection{Sensitivity analysis of the threshold $\mathcal{T}_0$}\label{App3}

The explicit representation~\eqref{eq:R0App} of the basic reproduction number $\mathcal{R}_0$ allows to study the sensitivity of $\mathcal{R}_0$ with respect to the several parameters of the model~\eqref{eq:SEIRwQ}. Moreover, thanks to Proposition~\ref{PropR0}, we know that the threshold
\begin{equation*}
\mathcal{T}_0 = \frac{\beta \tau [1 - (1 - r^2) p]}{(\rho\delta + \tau)(\sigma + \rho)} \left(
1 + \frac{\sigma\alpha}{\rho + \gamma_1}
\right)
\end{equation*}
can be used for an equivalent characterization of the condition $\mathcal{R}_0 < 1$. For this reason, it is handier to develop the sensitivity of $\mathcal{T}_0$ with respect to the parameters of the model, and deduce its dependence on perturbations of the parameters. We thus compute the normalized sensitivity index $S_x$ corresponding to the $x$ parameter, given by
$$
S_x := \frac{x}{\mathcal{T}_0}\frac{\partial\mathcal{T}_0}{\partial x}\, ,
$$
and we get that
\begin{align*}
    S_\beta & =  1 > 0\ ,\\[0.3ex]
    S_\tau & =  \frac{\rho\delta}{\rho\delta + \tau} > 0\ ,\\[0.3ex]
    S_p & = - \frac{(1-r^2)p}{1 - (1-r^2)p} < 0\ ,\\[0.3ex]
    S_r & = \frac{2r^2p}{1-(1-r^2)p} > 0\ ,\\[0.3ex]
    S_\delta & = - \frac{\rho\delta}{\rho\delta + \tau} < 0\ ,\\[0.3ex]
    S_\alpha & = \frac{\sigma\alpha}{\rho + \gamma_1 + \sigma\alpha} > 0\ ,\\[0.3ex]
    S_{\gamma_1} & = - \frac{\sigma\alpha\gamma_1}{(\rho + \gamma_1)(\rho + \gamma_1 + \sigma \alpha)} < 0\ ,\\[0.3ex]
     S_{\sigma} & = - \frac{\sigma[\gamma_1 + (1-\alpha)\rho]}{(\sigma + \rho)(\rho + \gamma_1 + \sigma \alpha)} < 0\ ,\\[0.3ex]
     S_{\rho} & = - \frac{\rho}{\rho + \gamma_1 + \sigma \alpha}\left[
     \frac{[\delta(\sigma + 2\rho)+\tau](\rho + \gamma_1 + \sigma\alpha)}{(\rho\delta + \tau)(\sigma + \rho)} + \frac{\sigma\alpha}{\rho + \gamma_1}
     \right] < 0\ .
\end{align*}
We thus notice the same qualitative dependence on the parameters already observed in Section~\ref{sec:R0}. In particular, if we increase $k$ times the parameter $\beta$, then $\mathcal{T}_0$ increases $k$ times as well. Similar deductions can be made on the other parameters, with the corresponding coefficients obtained by inserting the values of the parameters from Table~\ref{parameter_values}.
Moreover, from the expression of $S_\tau$ we realize that, if either $\rho$ or $\delta$ equal zero, then $\mathcal{T}_0$ does not depend on $\tau$ (as it happens for $\mathcal{R}_0$ as well, as noticed in Remark~\ref{rem:R0tau}). Similarly, if $\rho = 0$, then $S_\delta=0$, thus $\mathcal{T}_0$ does not depend on $\delta$. Regarding the parameters $p$ and $r$, their  dependence is mutually related as follows: if $p=0$, then $\mathcal{T}_0$ does not depend on $r$ (since $S_r = 0$), whereas if $r=1$ then $\mathcal{T}_0$ does not depend on $p$.

\bibliographystyle{plain}
\bibliography{covid}

\begin{thebibliography}{10}

\bibitem{An2020recovered}
J.~An, X.~Liao, T.~Xiao, S.~Qian, J.~Yuan, H.~Ye, F.~Qi, C.~Shen, Y.~Liu,
  L.~Wang, et~al.
\newblock {Clinical characteristics of the recovered COVID-19 patients with
  re-detectable positive RNA test}.
\newblock {\em medRxiv}, 2020.

\bibitem{backer2020incubation}
J.~A. Backer, D.~Klinkenberg, and J.~Wallinga.
\newblock {Incubation period of 2019 novel coronavirus (2019-nCoV) infections
  among travellers from Wuhan, China, 20--28 January 2020}.
\newblock {\em Eurosurveillance}, 25(5), 2020.

\bibitem{casella2020can}
F.~Casella.
\newblock {Can the COVID-19 epidemic be managed on the basis of daily data?}
\newblock {\em arXiv preprint arXiv:2003.06967}, 2020.

\bibitem{WHO-Ch}
WHO China.
\newblock {Report of the WHO-China Joint Mission on Coronavirus Disease 2019
  (COVID-19)}.
\newblock Technical report, World Health Organization,
  \url{https://www.who.int/docs/default-source/coronaviruse/who-china-joint-mission-on-covid-19-final-report.pdf},
  February 2020.

\bibitem{wiki:diamond}
Wikipedia contributors.
\newblock {COVID-19 pandemic on Diamond Princess}.
\newblock
  \url{https://en.wikipedia.org/wiki/COVID-19_pandemic_on_Diamond_Princess},
  2020.
\newblock Online; accessed.

\bibitem{Daym1375}
M.~Day.
\newblock Covid-19: four fifths of cases are asymptomatic, {C}hina figures
  indicate.
\newblock {\em BMJ}, 369, 2020.

\bibitem{diekmann1990definition}
O.~Diekmann, J.~A.~P. Heesterbeek, and J.~A.~J. Metz.
\newblock {On the definition and the computation of the basic reproduction
  ratio $\mathcal{R}_0$ in models for infectious diseases in heterogeneous
  populations}.
\newblock {\em J. Math. Biol.}, 28(4):365--382, 1990.

\bibitem{djidjou2020optimal}
R.~Djidjou-Demasse, Y.~Michalakis, M.~Choisy, M.~T. Sofonea, and S.~Alizon.
\newblock {Optimal COVID-19 epidemic control until vaccine deployment}.
\newblock {\em medRxiv}, 2020.

\bibitem{investigationgovuk}
Public~Health England.
\newblock {COVID-19: investigation and initial clinical management of possible
  cases}.
\newblock
  \url{https://www.gov.uk/government/publications/wuhan-novel-coronavirus-initial-investigation-of-possible-cases/priority-for-sars-cov-2-covid-19-testing},
  April 2020.

\bibitem{ferguson2020impact}
N.~M. Ferguson, D.~Laydon, G.~Nedjati-Gilani, N.~Imai, K.~Ainslie, M.~Baguelin,
  S.~Bhatia, A.~Boonyasiri, Z.~Cucunub{\'a}, G.~Cuomo-Dannenburg, et~al.
\newblock {Impact of non-pharmaceutical interventions (NPIs) to reduce COVID-19
  mortality and healthcare demand. Imperial College COVID-19 Response Team},
  2020.

\bibitem{guidanceCDC}
Centers for Disease~Control and Prevention.
\newblock {Evaluating and Testing Persons for Coronavirus Disease 2019
  (COVID-19)}.
\newblock
  \url{https://www.cdc.gov/coronavirus/2019-ncov/hcp/clinical-criteria.html},
  April 2020.

\bibitem{GrassPeseCecco2020}
G.~Grasselli, A.~Pesenti, and M.~Cecconi.
\newblock {Critical Care Utilization for the COVID-19 Outbreak in Lombardy,
  Italy: Early Experience and Forecast During an Emergency Response}.
\newblock {\em JAMA}, 323(16):1545--1546, 04 2020.

\bibitem{guardian2020}
The Guardian.
\newblock {Test, trace, contain: how South Korea flattened its coronavirus
  curve}.
\newblock
  \url{https://www.theguardian.com/world/2020/apr/23/test-trace-contain-how-south-korea-flattened-its-coronavirus-curve},
  April 2020.

\bibitem{OxCGRT}
T.~Hale, S.~Webster, A.~Petherick, T.~Phillips, B.~Kira, N.~Angrist, and
  Blavatnik~School of~Government.
\newblock {Coronavirus Government Response Tracker}.
\newblock
  \url{https://www.bsg.ox.ac.uk/research/research-projects/coronavirus-government-response-tracker},
  May 2020.

\bibitem{HarvardMedicine}
Harvard Medical~School Harvard Health~Publishing.
\newblock {If you've been exposed to the coronavirus}.
\newblock
  \url{https://www.health.harvard.edu/diseases-and-conditions/if-youve-been-exposed-to-the-coronavirus},
  March 2020.

\bibitem{Hu_etal_Asy}
Song C. Xu C. et~al. Hu, Z.
\newblock {Clinical characteristics of 24 asymptomatic infections with COVID-19
  screened among close contacts in Nanjing, China}.
\newblock {\em Sci. China Life Sci.}, 63(5):706--711, 2020.

\bibitem{aljazeeratimiline2020}
Al~Jazeera.
\newblock {Timeline: How the new coronavirus spread}.
\newblock
  \url{https://www.aljazeera.com/news/2020/01/timeline-china-coronavirus-spread-200126061554884.html},
  April 2020.

\bibitem{lauer2020incubation}
S.~A. Lauer, K.~H. Grantz, Q.~Bi, F.~K. Jones, Q.~Zheng, H.~R. Meredith, A.~S.
  Azman, N.~G. Reich, and J.~Lessler.
\newblock {The incubation period of coronavirus disease 2019 ({COVID}-19) from
  publicly reported confirmed cases: estimation and application}.
\newblock {\em Annals of Internal Medicine}, 2020.

\bibitem{lavezzo2020suppression}
E.~Lavezzo, E.~Franchin, C.~Ciavarella, G.~Cuomo-Dannenburg, L.~Barzon,
  C.~Del~Vecchio, L.~Rossi, R.~Manganelli, A.~Loregian, N.~Navarin, et~al.
\newblock {Suppression of COVID-19 outbreak in the municipality of Vo, Italy}.
\newblock {\em medRxiv}, 2020.

\bibitem{liang2020impacts}
J.~Liang and H.-Y. Yuan.
\newblock {The impacts of diagnostic capability and prevention measures on
  transmission dynamics of COVID-19 in Wuhan}.
\newblock {\em medRxiv}, 2020.

\bibitem{liu2020predicting}
Z.~Liu, P.~Magal, O.~Seydi, and G.~Webb.
\newblock {Predicting the cumulative number of cases for the COVID-19 epidemic
  in China from early data}.
\newblock {\em arXiv preprint arXiv:2002.12298}, 2020.

\bibitem{mizumoto2020estimating}
K.~Mizumoto, K.~Kagaya, A.~Zarebski, and G.~Chowell.
\newblock {Estimating the asymptomatic proportion of coronavirus disease 2019
  (COVID-19) cases on board the Diamond Princess cruise ship, Yokohama, Japan,
  2020}.
\newblock {\em Eurosurveillance}, 25(10):2000180, 2020.

\bibitem{outbreakncipcdcchina}
The~2019 nCoV Outbreak Joint Field Epidemiology Investigation~Team and Q.~Li.
\newblock {An Outbreak of NCIP (2019-nCoV) Infection in China - Wuhan, Hubei
  Province, 2019 - 2020}.
\newblock
  \url{http://weekly.chinacdc.cn/en/article/id/e3c63ca9-dedb-4fb6-9c1c-d057adb77b57},
  January 2020.

\bibitem{stephaniehegarty2020}
BBC News.
\newblock {The Chinese doctor who tried to warn others about coronavirus}.
\newblock \url{https://www.bbc.com/news/world-asia-china-51364382}, February
  2020.

\bibitem{nishiura2020estimation}
H.~Nishiura, T.~Kobayashi, T.~Miyama, A.~Suzuki, S.~Jung, K.~Hayashi,
  R.~Kinoshita, Y.~Yang, B.~Yuan, A.~R. Akhmetzhanov, et~al.
\newblock {Estimation of the asymptomatic ratio of novel coronavirus infections
  (COVID-19)}.
\newblock {\em medRxiv}, 2020.

\bibitem{whoreport392020}
World~Health Organization.
\newblock {Coronavirus disease 2019 (COVID-19)}.
\newblock Situation Report~39, World Heath Organization,
  \url{https://www.who.int/docs/default-source/coronaviruse/situation-reports/20200228-sitrep-39-covid-19.pdf?sfvrsn=5bbf3e7d_4},
  February 2020.

\bibitem{whoreport492020}
World~Health Organization.
\newblock {Coronavirus disease 2019 (COVID-19)}.
\newblock Situation Report~49, World Heath Organization,
  \url{https://www.who.int/docs/default-source/coronaviruse/situation-reports/20200309-sitrep-49-covid-19.pdf?sfvrsn=70dabe61_4},
  March 2020.

\bibitem{whoreport512020}
World~Health Organization.
\newblock {Coronavirus disease 2019 (COVID-19)}.
\newblock Situation Report~51, World Heath Organization,
  \url{https://www.who.int/docs/default-source/coronaviruse/situation-reports/20200311-sitrep-51-covid-19.pdf?sfvrsn=1ba62e57_10},
  March 2020.

\bibitem{whoreport331509}
World~Health Organization.
\newblock {Laboratory testing strategy recommendations for COVID-19}.
\newblock Interim guidance, World Heath Organization,
  \url{https://apps.who.int/iris/bitstream/handle/10665/331509/WHO-COVID-19-lab_testing-2020.1-eng.pdf},
  March 2020.

\bibitem{whopneumoniachina}
World~Health Organization.
\newblock {Pneumonia of unknown cause – China}.
\newblock
  \url{https://www.who.int/csr/don/05-january-2020-pneumonia-of-unkown-cause-china/en/},
  January 2020.

\bibitem{whocoronavirus2020}
World~Health Organization.
\newblock {Q}\&{A} on coronaviruses ({COVID-19}).
\newblock
  \url{https://www.who.int/emergencies/diseases/novel-coronavirus-2019/question-and-answers-hub/q-a-detail/q-a-coronaviruses},
  April 2020.

\bibitem{whoremarks16media2020}
World~Health Organization.
\newblock {WHO Director-General's opening remarks at the media briefing on
  COVID-19 - 16 March 2020}.
\newblock
  \url{https://www.who.int/dg/speeches/detail/who-director-general-s-opening-remarks-at-the-media-briefing-on-covid-19---16-march-2020},
  March 2020.

\bibitem{whoremarks3media2020}
World~Health Organization.
\newblock {WHO Director-General's opening remarks at the media briefing on
  COVID-19 - 3 March 2020}.
\newblock
  \url{https://www.who.int/dg/speeches/detail/who-director-general-s-opening-remarks-at-the-media-briefing-on-covid-19---3-march-2020},
  March 2020.

\bibitem{Read_etal}
J.~M. Read, J.~R.~E. Bridgen, D.~A.~T. Cummings, A.~Ho, and C.~P. Jewell.
\newblock {Novel coronavirus 2019-nCoV: early estimation of epidemiological
  parameters and epidemic predictions}.
\newblock {\em medRxiv}, 2020.

\bibitem{Remuzzi2020}
A.~Remuzzi and G.~Remuzzi.
\newblock {COVID-19 and Italy: what next?}
\newblock {\em The Lancet, Health Policy}, 395:1225--28, 2020.

\bibitem{shen2020modelling}
M.~Shen, Z.~Peng, Y.~Xiao, and L.~Zhang.
\newblock {Modelling the epidemic trend of the 2019 novel coronavirus outbreak
  in China}.
\newblock {\em bioRxiv}, 2020.

\bibitem{shi2020seir}
P.~Shi, S.~Cao, and P.~Feng.
\newblock {SEIR Transmission dynamics model of 2019 nCoV coronavirus with
  considering the weak infectious ability and changes in latency duration}.
\newblock {\em medRxiv}, 2020.

\bibitem{time-reinfection}
Time.
\newblock {Can You Be Re-Infected After Recovering From Coronavirus? Here's
  What We Know About COVID-19 Immunity}.
\newblock \url{https://time.com/5810454/coronavirus-immunity-reinfection/},
  April 2020.

\bibitem{newyorkqina2020}
The New~York Times.
\newblock {China Reports First Death From New Virus}.
\newblock
  \url{https://www.nytimes.com/2020/01/10/world/asia/china-virus-wuhan-death.html},
  January 2020.

\bibitem{vdDW02}
P.~van~den Driessche and J.~Watmough.
\newblock {Reproduction numbers and sub-threshold endemic equilibria for
  compartmental models of disease transmission}.
\newblock {\em Mathematical Biosciences}, 180:29--48, 2002.

\bibitem{wajnberg2020humoral}
A.~Wajnberg, M.~Mansour, E.~Leven, N.~M. Bouvier, G.~Patel, A.~Firpo, R.~Mendu,
  J.~Jhang, S.~Arinsburg, M.~Gitman, et~al.
\newblock Humoral immune response and prolonged pcr positivity in a cohort of
  1343 sars-cov 2 patients in the new york city region.
\newblock {\em medRxiv}, 2020.

\bibitem{wang2020updated}
W.~Wang, J.~Tang, and F.~Wei.
\newblock {Updated understanding of the outbreak of 2019 novel coronavirus
  (2019-nCoV) in Wuhan, China}.
\newblock {\em Journal of Medical Virology}, 92(4):441--447, 2020.

\bibitem{Woelfel_etal}
R.~Woelfel, V.~M. Corman, W.~Guggemos, M.~Seilmaier, S.~Zange, M.~A. Mueller,
  D.~Niemeyer, P.~Vollmar, C.~Rothe, M.~Hoelscher, T.~Bleicker, S.~Bruenink,
  J.~Schneider, R.~Ehmann, K.~Zwirglmaier, C.~Drosten, and C.~Wendtner.
\newblock {Clinical presentation and virological assessment of hospitalized
  cases of coronavirus disease 2019 in a travel-associated transmission
  cluster}.
\newblock {\em medRxiv}, 2020.

\bibitem{worldometers-us}
Worldometer.
\newblock {Coronavirus cases in US}.
\newblock \url{https://www.worldometers.info/coronavirus/country/us/}, 2020.

\bibitem{worldometers-testing}
Worldometer.
\newblock {Coronavirus Testing: Criteria and Numbers by Country}.
\newblock \url{https://www.worldometers.info/coronavirus/covid-19-testing/},
  2020.

\bibitem{worldometers-countries}
Worldometer.
\newblock {Reported Cases and Deaths by Country, Territory, or Conveyance}.
\newblock \url{https://www.worldometers.info/coronavirus/#countries}, 2020.

\bibitem{newyork-missingdeaths}
J.~Wu, A.~McCann, J.~Katz, and E.~Peltier.
\newblock {73,000 Missing Deaths: Tracking the True Toll of the Coronavirus
  Outbreak}.
\newblock
  \url{https://www.nytimes.com/interactive/2020/04/21/world/coronavirus-missing-deaths.html},
  May 2020.

\bibitem{zhang2020evolving}
J.~Zhang, M.~Litvinova, W.~Wang, Y.~Wang, X.~Deng, X.~Chen, M.~Li, W.~Zheng,
  L.~Yi, X.~Chen, et~al.
\newblock {Evolving epidemiology and transmission dynamics of coronavirus
  disease 2019 outside Hubei province, China: a descriptive and modelling
  study}.
\newblock {\em The Lancet Infectious Diseases}, 2020.

\bibitem{Zhou2020}
F.~Zhou, T.~Yu, R.~Du, G.~Fan, Y.~Liu, Z.~Liu, J.~Xiang, Y.~Wang, B.~Song,
  X.~Gu, et~al.
\newblock {Clinical course and risk factors for mortality of adult inpatients
  with COVID-19 in Wuhan, China: a retrospective cohort study}.
\newblock {\em The Lancet}, 2020.

\end{thebibliography}

\end{document}